\documentclass[final,10pt,twocolumn]{elsarticle}

\usepackage[top=2.5cm,bottom=2.5cm,left=2.5cm,right=2.5cm]{geometry}

\usepackage{hyperref}
\usepackage{multirow}
\usepackage{amsmath,amsfonts,amssymb,amscd,amsthm,xspace}
\usepackage{graphicx}
\usepackage{epstopdf}
\usepackage{booktabs}
\usepackage[shortlabels]{enumitem}
\usepackage[usenames,dvipsnames,svgnames,table]{xcolor}
\usepackage{rotating}
\usepackage{listings}
\usepackage{longtable}

\journal{Nuclear Instrument and Methods in Physics Research A}

\begin{document}

\begin{frontmatter}

\title{The Borexino Thermal Monitoring \& Management System and simulations of the fluid-dynamics of the Borexino detector under asymmetrical, changing boundary conditions}

\author[myamericanaddress,myitalianaddress]{D. Bravo-Bergu\~no}

\author[RiccardoAddress]{R. Mereu}
\author[myamericanaddress]{P. Cavalcante}
\author[LNGS]{M. Carlini}
\author[Andrea]{A. Ianni}
\author[Andrea]{A. Goretti}
\author[LNGS]{F. Gabriele}
\author[myamericanaddress]{T. Wright}
\author[myamericanaddress]{Z. Yokley}
\author[myamericanaddress]{R.B. Vogelaar}
\author[Andrea]{F. Calaprice}
\author[RiccardoAddress]{F. Inzoli}

\address[myamericanaddress]{Physics Department, Virginia Tech, 24061 Blacksburg, VA (USA)}
\address[myitalianaddress]{INFN Sezione Milano, Via Celoria 16, 20133 Milano (Italy) - (+39) 3394636849 - \textit{david.bravo@mi.infn.it}}
\address[RiccardoAddress]{Department of Energy - Politecnico di Milano, via Lambruschini 4, 20156 - Milano, Italy}
\address[Andrea]{Physics Department, Princeton University, Princeton, NJ 08544, USA}
\address[LNGS]{INFN Laboratori Nazionali del Gran Sasso, 67010 Assergi (AQ), Italy}

\begin{abstract}
A comprehensive monitoring system for the thermal environment inside the Borexino neutrino detector was developed and installed in order to reduce uncertainties in determining temperatures throughout the detector. A complementary thermal management system limits undesirable thermal couplings between the environment and Borexino's active sections. This strategy is bringing improved radioactive background conditions to the region of interest for the physics signal thanks to reduced fluid mixing induced in the liquid scintillator. Although fluid-dynamical equilibrium has not yet been fully reached, and thermal fine-tuning is possible, the system has proven extremely effective at stabilizing the detector's thermal conditions while offering precise insights into its mechanisms of internal thermal transport. Furthermore, a Computational Fluid-Dynamics analysis has been performed, based on the empirical measurements provided by the thermal monitoring system, and providing information into present and future thermal trends. A two-dimensional modeling approach was implemented in order to achieve a proper understanding of the thermal and fluid-dynamics in Borexino. It was optimized for different regions and periods of interest, focusing on the most critical effects that were identified as influencing background concentrations. Literature experimental case studies were reproduced to benchmark the method and settings, and a Borexino-specific benchmark was implemented in order to validate the modeling approach for thermal transport. Finally, fully-convective models were applied to understand general and specific fluid motions impacting the detector's Active Volume.
\end{abstract}


\end{frontmatter}


\section{Introduction}
\label{sec:intro}

The Borexino liquid scintillator (LS) neutrino observatory is devoted to performing high-precision neutrino observations, and is optimized for measurements in the low energy (sub-MeV) region of the solar neutrino spectrum. Borexino has succeeded in determining all major solar neutrino flux components already with its first dataset \textit{Phase 1} (2007-10): first direct detections of \textit{pp}\cite{pp}, \textit{pep}\cite{pep} and $^7$Be\cite{7Be} neutrinos; and lowest-threshold observation of $^8$B\cite{8B} at 3 MeV, as well as the best available limit in the CNO solar $\nu$ flux\cite{8B}. More recently, high-precision (down to $\approx$2.8$\%$) determinations of the aforementioned solar neutrino fluxes have been attained using new techniques and enlarged statistics from the post-LS-purification phase: \textit{Phase 2}\cite{wideband}\cite{new8B}. Geoneutrinos have also been measured with high significance (5.9$\sigma$\cite{geo}) by Borexino, thanks to the extremely clean $\overline{\nu}_e$ channel -- which is expected to gain even more relevance during the \textit{Short-distance neutrino Oscillations with boreXino} (SOX) phase of the experiment. An $\overline{\nu}_e$ generator made of $^{144}$Ce-Pr will be placed in close proximity to the detector during CeSOX, in order to probe for anomalous oscillatory behaviors and unambiguously check for experimental signatures along the phase space light sterile neutrinos might lie in\cite{SOX} \cite{Giunti}. These results are possible thanks to the unprecedented, extremely radio-pure conditions reached in the Active Volume (AV) of the detector (down to $\leq$10$^{-19}$ g($^{239}$U/Th)/g(LS)\cite{wideband}) -- achieved thanks to a combination of ultra-clean construction and fluid-handling techniques, as well as dedicated scintillator purification campaigns\cite{purif}. Detailed detector response determination was made possible thanks to very successful internal calibration campaigns\cite{calib} which did not disturb the uniquely radio-pure environment. 

The unprecedented radiopurity levels reached in Borexino's LS are the key to the uniqueness of the detector's results. The conditions reached for \textit{Phase 2} after the purification campaign in 2011\cite{wideband} have raised the need for increased stability in their spatio-temporal distribution inside the detector. Indeed, mixing of the free scintillating fluid in the AV could cause unwanted background fluctuations that, with careful management measures, may be minimized or avoided by means of external thermal environment control and stabilization. It is assumed the extremely dilute concentrations of background radioisotopes are carried by the fluid movement in ideally point-like, non-interacting particulates Ðthereby establishing a direct correlation between fluid dynamics and background migration, which is only attenuated through the corresponding halflives.

Section~\ref{sec:stability} of this paper will detail the correlation existent between background stability and detector thermal conditions. Section~\ref{sec:monitoring} will deal with the temperature monitoring solution devised and installed in Borexino in order to inform and manage the deployment of the management systems discussed in Section~\ref{sec:management}. Together, this hardware is referred to as the \textit{Borexino Thermal Monitoring and Management System} (BTMMS). Section~\ref{sec:exp_results} highlights the experimental results obtained by the BTMMS. Section~\ref{sec:conductive} will focus on the conductive Computational Fluid Dynamics (CFD) simulations developed in order to more comprehensively understand the detector past and future thermal behavior. Section~\ref{sec:methodology} will detail the technical premises that baseline the convective simulation strategy, while Section~\ref{sec:benchmarking} will discuss the benchmarking strategy for the CFD convective simulations that would yield confidence on the simulative approach used, as well as the empirical data-based thermal benchmarking performed for the Borexino geometry. Section~\ref{sec:SSS_IV} will instead focus on the Inner Detector models developed in order to understand fluid movement inside this closed, near-equilibrium system. It will also showcase the evolution toward a more focused model with greater detail around Borexino's Active Volume (AV). Finally, Section~\ref{sec:conclusions} will give a comprehensive discussion of the conclusions reached through these models, their impact in future detector operations and physics results, and the perspectives on new studies building upon the present work.

\section{Background stability}
\label{sec:stability}

\begin{figure}[ht]
\centering\small\includegraphics[width=1\linewidth]{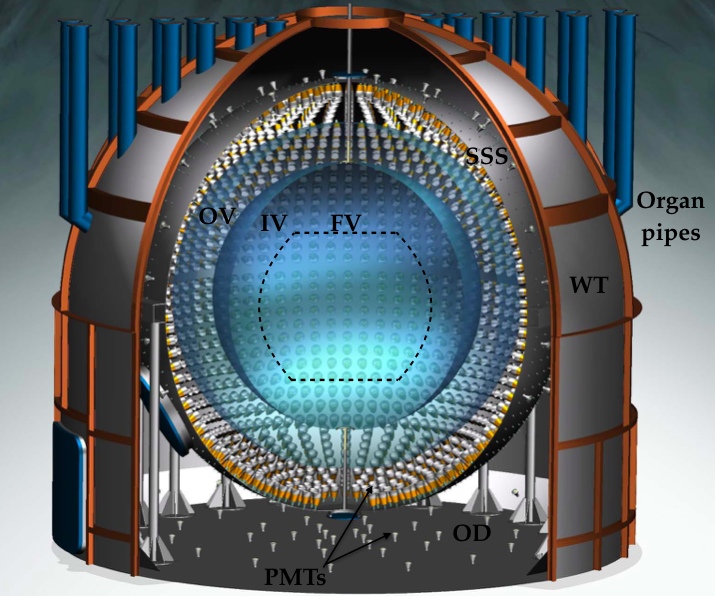}
\caption{The Borexino neutrino observatory, with its main structures annotated. See full text for details.}
\label{fig:BX}
\end{figure}

Borexino, located in the Hall C of the Gran Sasso National Laboratories' (LNGS) underground facilities (3,800 m w.e.), measures solar neutrinos via their interactions with a 278 tonnes target of organic liquid scintillator. The ultrapure liquid scintillator, pseudocumene (1,2,4-trimethylbenzene, PC) solvent with 1.5 g/l of 2,5-diphenyloxazole (PPO) scintillating solute, is contained inside a thin transparent spherical nylon Inner Vessel (IV) of 8.5 m diameter. Solar neutrinos are detected by measuring the energy and position of electrons scattered by neutrino-electron elastic interactions. The scintillator promptly converts the kinetic energy of electrons by emitting photons, which are detected and converted into electronic signals, \textit{i.e.} photoelectrons (p.e.), by 2,212 photomultipliers (PMTs) mounted on a concentric 13.7 m-diameter stainless steel sphere (SSS, see Figure~\ref{fig:BX}). A software-defined, analysis-dependent Fiducial Volume (FV) is established inside the IV. The volume between the nylon vessel and the SSS is filled with 889 tonnes of ultra pure, non scintillating fluid called "buffer" acting as a radiation shield for external gamma rays and neutrons. A second, larger nylon sphere (Outer Vessel (OV), 11.5 m diameter) prevents radon and other radioactive contaminants from the PMTs and SSS from diffusing into the central sensitive volume of the detector, and segments the Inner and Outer Buffers (IB and OB). The SSS is immersed in a 2,100-tonne Water Tank (WT) acting as a \v{C}erenkov detector tagging residual cosmic $\mu^{\pm}$.

Radioactive decays within the scintillator form a background that can mimic neutrino signals. A record low scintillator contamination of $\leq$10$^{-19}$ g/g was achieved for $^{238}$U and $^{232}$Th. 

\begin{figure}[!htb]
\centering
\minipage{0.46\textwidth}
  \includegraphics[width=\linewidth]{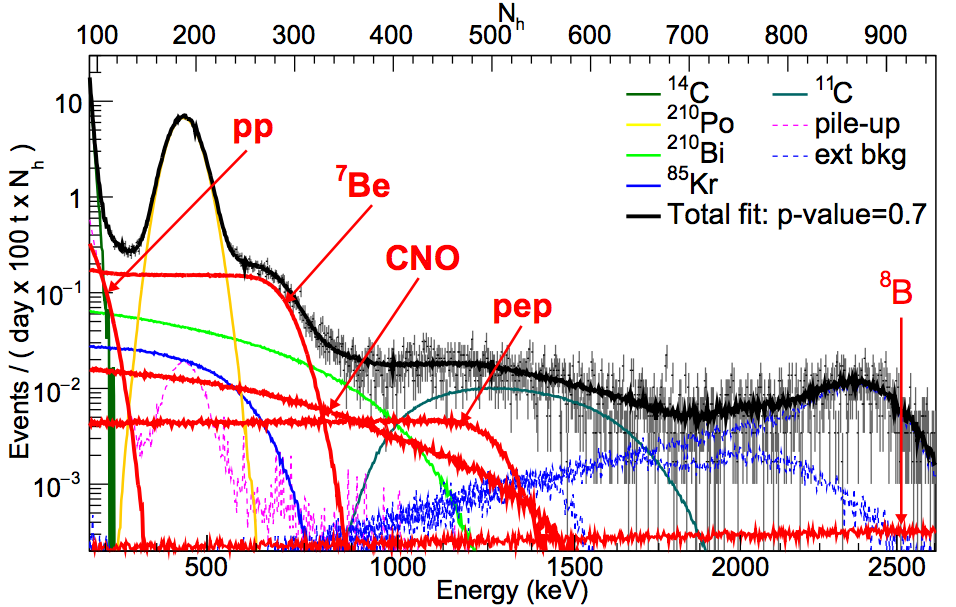}
  \caption{\small{Borexino spectrum in the main analysis energy range ($\approx$500 p.e./MeV), from \cite{wideband}. Note the near-degeneracy in the $\approx$850 keV area between the spectral shapes of $^{210}$Bi and CNO $\nu$, the only window where $\nu_{CNO}$ are prevalent over $pp$-chain $\nu$. Color version available online.}}\label{fig:spectrum}
\endminipage\hfill
\minipage{0.48\textwidth}
  \includegraphics[width=\linewidth]{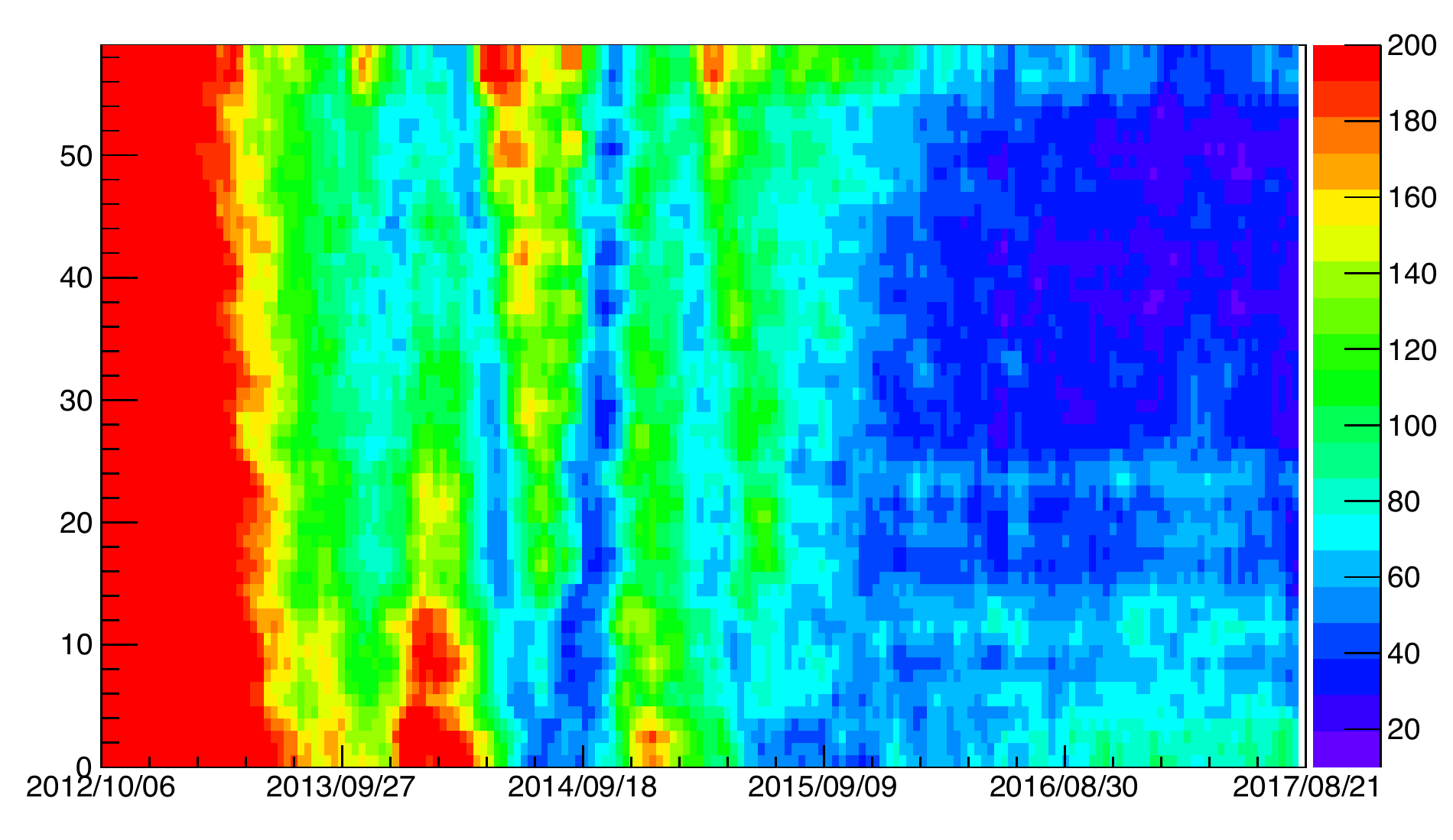}
  \caption{\small{Regional map of polonium concentrations with respect to time (color code: cpd/100 tonnes). Expected plateau levels due to equilibrium $^{210}$Bi in the scintillator are $\approx$20 cpd/100 tonnes. The Y axis indicates regional subdivisions of a 3m-radius FV, running from its bottom to its top. Color version available online.}}\label{fig:polonium}
\endminipage
\end{figure}

Of particular importance to the future of Borexino is the effort toward measuring the sub-leading, but crucial, CNO solar neutrino component ($\leq$1$\%$ of the Sun's output\cite{CNO}). Its neutrino recoil spectral shape (endpoint at $E_{max}$=1.74 MeV) places it under several intrinsic Borexino backgrounds, in particular $^{85}$Kr and $^{210}$Bi, whose spectral shapes exhibit a large degree of correlation with the solar neutrino signal -- especially so for $^{210}$Bi in the $\approx$400 p.e. energy window, where the CNO rate is expected to be higher than the neighboring $^7$Be and $pep$ neutrinos (see Figure~\ref{fig:spectrum}). It is estimated a $\lesssim$10$\%$ precision in the determination of the $^{210}$Bi concentration in Borexino's FV is needed, during a long enough time period, to collect the very low expected CNO $\nu$ counts ($\approx$3-5 cpd/100 tonnes).

Its decay daughter $^{210}$Po provides an accurate method for succeeding in this determination. Indeed, $^{210}$Po's $\alpha$ decay allows for it to be efficiently tagged out through Pulse-Shape Discrimination (PSD) techniques with very low inefficiencies. Conversely, the $\beta ^-$ decay of bismuth (Q=1160 keV, $t_{1/2}$=5 days) provides an indistinguishable (only statistically-subtractable) signal to $\nu_e-e^-$ elastic scatterings which cannot determine the rate down to the required uncertainty levels, due to the shape degeneracy between the bismuth and solar $\nu$ components in the so-called "bismuth valley"\footnote{The "bismuth valley" is the energy window where CNO $\nu$s are least overwhelmed by other solar neutrinos or irreducible background components, between the $^7$Be shoulder and $^{11}$C+$\nu_{pep}$, around 400 p.e..}. Indeed, once initial out-of-equilibrium $^{210}$Po levels have decayed away (initial rate $\approx$800 times higher than that of bismuth; $t_{1/2}$=138.4 days), the decay curve would asymptotically reach a plateau baseline corresponding to the secular equilibrium levels of $^{210}$Bi. This condition has been close at hand for most of Borexino's \textit{Phase 2} DAQ period -- but new out-of-equilibrium, regionally-significant fluctuations in the $^{210}$Po levels have prevented reaching it (see Figure~\ref{fig:polonium}). A dedicated publication on the specialized signal analysis for $^{210}$Bi-Po and its interpretation is in preparation.

Crucially, it is known $^{210}$Pb (parent of $^{210}$Bi, $t_{1/2}$=22.3 years, off-threshold-low Q-value) exhibits higher concentrations in the IV, and consequently provides a continuous, "inexhaustible" source of $^{210}$Bi-Po. Historically, $^{210}$Po fluctuations show a correlation with large environmental temperature excursions in the experimental Hall, hinting at a possible mechanism for replenishment of out-of-equilibrium polonium in the FV: fluid mixing through temperature-driven convection from the AV's periphery, around the IV, toward the center. Additionally, the regional homogeneity and stability in $^{210}$Bi concentration suggests that the underlying fluid-dynamics are slow enough to prevent most of this isotope to be transported inside the FV faster than it decays, establishing a soft upper limit in radial fluid velocity of $\frac{\mathcal{O}(m)}{\mathcal{O}(2\cdot10^6 s)}\lesssim$5$\cdot$10$^{-7}$ m/s.

A yearly modulation in the asymmetry between $^{210}$Po concentration at the top and the bottom of the FV started being apparent in 2013, suggesting a correlation with external temperature trends. No azimuthal dependence is apparent, and the expected direct proportionality dependence is verified in the radial direction. Borexino's legacy thermal monitoring system, installed during the detector's construction, could provide readings on its vertical axis (8 sensors in IB and OB) and SSS surface (28 sensors), as well as selected points in the environmental air around it. They also showed the correlation between ambient temperature upsets in Hall C and background concentration spikes. However, their age and design requirements made them not the ideal system to monitor changes influencing fluid-dynamics in the IV, owing to signal coarseness, sensor position and irrecoverability for replacement or recalibration. The needed system would allow for a fine mapping of Borexino's temperature profile, as close as possible to the IV but also monitoring the thermal transport from the environment. Additionally, by determining the "ideal" temperature profile, the stabilization of the detector's thermal distribution could be attempted.

\section{Monitoring system}
\label{sec:monitoring}

Borexino's natural temperature profile exhibits a stable stratification based on a temperature gradient that increases monotonically with height. Denser isotherms are present in the bottom half, indicating a sharper gradient in that region, that then smooths out toward the top, where temperatures are more uniform. Heat is exchanged with the rock, steel and concrete bottom, as well as with the air surrounding the WT, and ideally should contribute to keeping that gradient profile. In reality, local temperature inhomogeneities in the ambient air as well as seasonal upsets, deviate from this ideal situation, and generate spatial and temporal perturbations. For this reason, the Latitudinal Temperature Probe System (LTPS) is conceived as a vertical profile monitoring system, with an azimuthal resolution of 180$^{\circ}$.
\paragraph{\textbf{Layout}} The LTPS consists of three subsystems (see conceptual diagram in Figure~\ref{fig:positions}): 
\begin{enumerate}
\item \textit{Phase I}: 28 internal probes, located in the \textit{re-entrant tube} (RET) system. These ports were in principle envisioned for the insertion of small, low-activity sources for PMT and Outer Detector calibration\cite{ext_calib}, but are otherwise empty. They are located next to the PMT cable ports ("organ pipes") on top of Borexino, allowing for direct access to the SSS, up to $\approx$0.5 m into the OB.
\item \textit{Phase II}: 20 external probes located on the WT's outer wall surface, complemented by 6 probes located in a T-shaped service tunnel (1 m$^2$ section) under the detector (\textit{SOX pit}).
\item \textit{Phase III}: 6 external probes located on the WT's upper dome, one inside the calibration clean-room (CR4) located over Borexino's vertical axis, as well as several probes for exterior ambient air readings.
\end{enumerate}
The LTPS probes (see technical specifications in Table~\ref{table:Specifications}) output a voltage differential that is routed through the Signal Conditioning Box (SCB), to then be routed to the LabQuest Mini 12-bit digitizer outputting the raw data for each probe (integer number scaled to 16 bits). Once sent to the interface computer, a \texttt{C++} program converts the raw data back to voltages and temperatures according to empirically-determined calibration functions. 

\begin{figure}[ht]
\centering\small\includegraphics[width=1\linewidth]{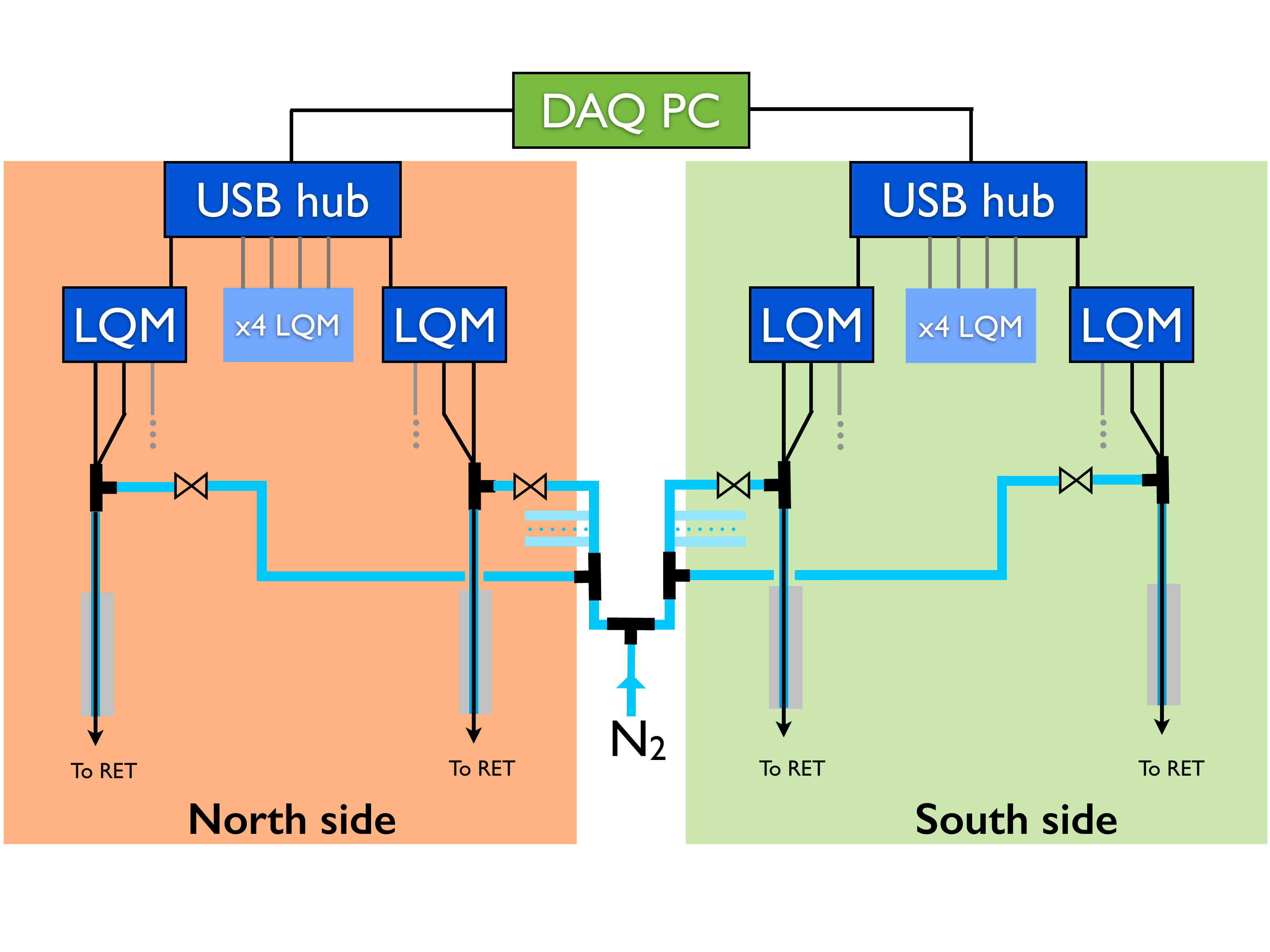}
\caption{Scheme for the 28 sensors in the LTPS Phase I layout, including electrical and gas connections. LQMs are the electronics interfaces supporting 3 sensors, and as shown 6 of them are present on each side. Phase II.a shows a similar design with just one sensor per sheathing and no purging capability. Phase II.b and III have simpler routings with no N/S sides due to their special configurations. RET stands for "Re-Entrant Tubes", or the ports that reach 0.5 m inside the SSS into the OB. Color version available online.}
\label{fig:LTPS_scheme}
\end{figure}

Phase I internal probes are sheathed inside a low-friction PVC tube (10 mm OD, 8 mm ID) terminated with a small section (8 mm OD) polyethylene tube providing support for the sensor tip and avoiding disconnection between it and the sheathing. A purging and drying nitrogen flux is facilitated by slits cut at the front end of the sheathings, and fed through a manifold through their top end, in order to clear out small amounts of condensation in the ports that could damage the probes in the long run. Because of this design (see Figure~\ref{fig:LTPS_scheme}), the internal sensors are easily accessible for removal, replacement or relocation. 

\begin{figure}[ht]
\centering\small\includegraphics[width=0.7\linewidth]{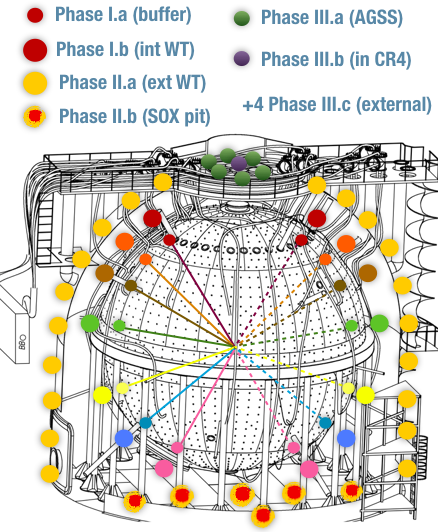}
\caption{Conceptual design of LTPS sensor positions within Borexino. Although shown to lie on the same plane, Phase I and II sensors actually show some scatter around $\phi$=0 to avoid SSS structures.}
\label{fig:positions}
\end{figure}

The internal probes are subdivided in Phase I.a and I.b, corresponding to the probes measuring OB (0.5 m inside the SSS) and WT (0.5 m outside the SSS) temperatures, respectively. Being separated by a meter, they share the same sheathing. Phase I.a started data-taking on October 29th, 2014 and Phase I.b on April 10th, 2015.

\begin{table}[ht]
\centering
\begin{tabular}{l l}
\hline
\textbf{Specification} & \textbf{Value} \\
\hline
\textit{Order Code} & TPL-BTA \\
\textit{Temp. transducer model} & AD590JH \\
\textit{Cable length} & 30 m \\
\textit{Maximum diameter} & 7 mm\\
\textit{Range} & -50$^{\circ}$C -- 150$^{\circ}$C  \\
\textit{Specified accuracy} & $^+_-$0.2$^{\circ}$C  \\
\textit{Specified resolution} & 0.07$^{\circ}$C \\
\textit{Power} & 7.4 mA at 5 VDC \\
\textit{Response time} & 8-10 s (still water)\\
 & 45 s (stirred water) \\
 & 100 s (moving air) \\
\hline
\end{tabular}
\caption{Vernier Extra-Long Probes specifications.}
\label{table:Specifications}
\end{table}

Phase II.a and III.a sensors are located inside flexible tubing at fixed locations, in order to facilitate access and replacement as for Phase I sensors, and lie under the layers of the insulation system described in Section~\ref{sec:management}, and measure the external WT wall boundary condition. They went online between summer 2015 and early 2016, depending on their position. Phase II.b probes in the tunnel measure the bottom heat sink boundary condition, both at the tunnel ceiling (Borexino's bottom) as well as the absolute acquifer-controlled rock temperature on the pit floor ($\approx$6.5$^{\circ}$C) and started data-taking in autumn 2015. 

\paragraph{\textbf{Calibration}} Given the level of precision needed, a custom calibration was performed in order to improve the probes' factory specifications. This had the objectives of characterizing the probes' behavior and detect eventual individual anomalous outputs, check their short- and long-term stability, and further calibrate them in order to minimize systematic uncertainties by adding individually-tailored calibration coefficients (offset and -- if necessary -- a linear term), and fine-tuning of their correction factors to an absolute reference temperature. This was achieved, in this order, with (i) characterization runs in air (low precision), (ii) absolute temperature baths (0$^{\circ}$C and ambient temperature), (iii) relative benchmarking of all probes in a single re-entrant port with constant temperature within the calibration time, for relative probe-to-probe correction factor tweaking, and (iv) absolute water bath cross-check at nominal working temperature ranges. Only the 14 Phase I.a sensors used all these techniques -- lessons learned with them were extrapolated to the rest, which were only calibrated using (iv). This also avoided downtime in the already-installed Phase I.a LTPS, which would otherwise be introduced with (iii).

The calibrations yielded a $\leq$0.04$^{\circ}$C relative accuracy and a similar level of absolute precision in LTPS Phase I. Phase II and III sensors, not having been as exhaustively cross-calibrated, would exhibit a slightly worse absolute precision, but their more peripheral position renders this uncertainty negligible.
\paragraph{\textbf{DAQ Software}} Data processing in each of the 3 readout computers is based on a custom-modified version of Vernier's \texttt{NGIO\_DeviceCheck} simple acquisition code with a typical 30-minute refresh time. It is then stored in a PSQL database. Online and offline visualization and analysis tools were developed for each LPTS Phase.
\paragraph{\textbf{Technical specs results}} The LTPS sensors showed increased performance with $\approx$20$\%$ of the intrinsic jitter of the old, legacy Borexino internal thermometers ($\approx$0.01$^{\circ}$C). Long-term stability of $\leq$0.05$^{\circ}$C was also observed, in multi-hour-long periods. With these characteristics and the relative position of the Phase I.a and I.b internal sensors (separated by a distance of $\approx$1 m and measuring temperatures across the SSS' inner and outer walls) thermal transport studies to characterize the thermal transport latency could be performed. Through functional fitting of relevant transient features seen in the temporal evolution plot, the time delay between features could be quantitatively measured. An upper limit constraint of 18-24 h/m (i.e., 1-1.3 m/day) for the thermal transport speed across the SSS boundary was determined. This demonstrates a low thermal inertia to detector-wide temperature transients: extrapolating this to the bulk of the detector, even minor thermal transients would reach the IV in around 4 days.

Another relevant feature is the existence of a slight, but pervasive thermal asymmetry between the North and South sides of the detector, assumed to be mostly due to the different external environment each side is subjected to. The CFD sections of this work show the importance of these temperature asymmetries in detector fluid dynamics. These asymmetries depend on latitude, while the gradient oscillates every few months, being alternatively slightly larger on the North or on the South. From the summer of 2016, the gradient has a pronounced asymmetry, with the North being $\approx$0.1$^{\circ}$C colder than the South. Due to environmental conditions, the North side also exhibits a "hot spot" on its lowermost Phase II.a sensor (close to the ground level of the WT wall), but this is a very localized effect with no widespread importance, and has been shown not to affect the IV in any major way.

An additional, accidental blind check of the validity of the cross-calibration performed for the Phase I probes was offered by a slight mismatch between the nominal positions of the re-entrant tubes and the actual ones: some of them were slightly displaced due to structural interference with other SSS hardware. This was most noticeable in the equatorial sensors (+7$^{\circ}$ nominal latitude), which exhibit a 3$^{\circ}$ difference, with the South sensor being lower than the North one -- and therefore, with the former showing a $\approx$-0.2$^{\circ}$C chronic unexplained offset, until the positioning issue was clarified by cross-checks with the external calibration reconstructed source positions\cite{ext_calib}.

\section{Management system}
\label{sec:management}

The Thermal Insulation System (TIS) was installed in parallel to the Phase II.a probes in the May-December 2015 timeframe, with the aim of reducing the temperature transients clearly seen in data from the Phase I probes, effectively increasing the thermal resistance of Borexino's largest boundary: the WT skin. It consists of a double layer of mineral wool material (Ultimate Tech Roll 2.0 - \textit{Isover}) that covers the full surface to a depth of 20 cm. In the thermal region of interest for Borexino, it has an extremely low conductivity value of $\approx$0.03-0.04 W/m$\cdot$K (see Table~\ref{table:isover}).

\begin{table}[ht]
\centering
\begin{tabular}{l c c c}
\textbf{Characteristics} & \textbf{Value} & \textbf{Units} & \textbf{EN std.} \\
\hline
Fire class & A1 & - & 13501-1 \\
Max temperature & 360 & $^{\circ}$C & 14706 \\
($>$250 Pa) & & \\
Airflow resistivity & 10 & kPa$\cdot$s/m$^2$ & 29053 \\
Acoustic absor. & 0.81 & $\alpha_w$ & ISO11654 \\
\textit{$\lambda_D$} (10$^{\circ}$C) & 0.033 & W/m$\cdot$K & 12667 \\
\textit{$\lambda_D$} (50$^{\circ}$C) & 0.040 & W/m$\cdot$K & 12667 \\
\textit{$\lambda_D$} (100$^{\circ}$C) & 0.050 & W/m$\cdot$K & 12667 \\
\end{tabular}
\caption{Technical specifications for the TIS thermal insulation material Ultimate Tech Roll 2.0 (Isover). $\lambda_D$ stands for thermal conductivity.}
\label{table:isover}
\end{table}

The exterior layer features a reflective aluminized film reinforced with an internal fiber glass grid, as well as a metallic wire mesh netting on the outside face (Ultimate Protect Wired Mat 4.0 Aluminized Isover). Metallic anchors (20 cm long) were epoxyed on the WT walls in order to support the TIS sections, with a surface density of $\approx$5/m$^2$. An approximate $>$1000 m$^2$ of detector surface were insulated, including the "organ pipes" through which the PMT cables enter the tank toward the SSS, and the interior floor of the calibration cleanroom located on the top dome. An extra $\approx$430 m$^2$ of I-beam structural elements' surface was also insulated, with just the 10 cm-thick aluminized insulation layer. Insulation started by the main surfaces, from the bottom up. Once the horizontal walls were approximately covered, I-beams followed, together with the dome's lower "rings" and largest organ pipe sections.

The Active Gradient Stabilization System (AGSS) was installed on the uppermost dome "ring" section surrounding the cusp-mounted calibration clean-room, before covering the metal wall with the insulation. It was conceptualized in order to avoid possible transient or long-term effects negatively affecting fluid stability, through the maximization of a positive thermal gradient between top and bottom of the detector, as well as the minimization of external disturbances in the top of Borexino's WT dome: the most vulnerable boundary area to environmental air temperature upsets.

The AGSS consists of twelve $\approx$18 m-long independent water loop circuits, based on 14 mm-OD copper serpentine tubing. The transfer from the circuits to the 12 input/output manifold occurs through a multilayered insulated pipe in order to guarantee a satisfactory thermal decoupling from the environmental temperature. A 3 m$^3$/h centrifugally-pumped water heater (3 kW) provides the heating power, controlled by a $\pm$0.1$^{\circ}$C-accurate controller. Nevertheless, even at constant temperature, the AGSS heating would provide stabilization, since its surface is quite small compared to the total surface of the WT dome, and it would provide a thermal anchoring effect on the topmost fluid, even if causing weak local convection in just the topmost water.  The system also includes an expansion tank, a mass flowmeter to manually adjust the flow and several safety items (mechanical thermostat, safety valve, pressure and flow switches). Furthermore, maximal bonding to the subjacent WT dome is ensured through copper anchors and a layer or aluminized tape, to ensure directional heat transfer toward the bottom of the heat exchanger assembly. The serpentines are horizontally distributed, in order to allow the most heated water to enter it on the upper inlet, and exit it through the lowermost outlet, ensuring maximal heat transfer at the top of the WT and keeping it constant (or reduced) toward the bottom. 

A slow control system was implemented with data readout from 12 thermocouple precision temperature probes, in conjunction with the pressure and flow switches readouts. Constant power and constant temperature modes are available. Six of those are LTPS sensors (Phase III.a) interleaved with the serpentine in order to provide context data, while 4 outlet and 2 inlet separate sensors are used for heat transfer calculation. AGSS operation was started on January 10th, 2017, with a setpoint equal to the dome's water temperature. From late January 2017, the setpoint was raised by 0.1$^{\circ}$C/week to test the system, and by late summer 2017 it had reached a setpoint neighboring the maximum local aestival temperature. Efforts to better control the Hall C's environmental temperature are also underway on the Laboratories' side.

\section{Experimental results}
\label{sec:exp_results}

\begin{figure}[ht]
\centering\small\includegraphics[width=1\linewidth]{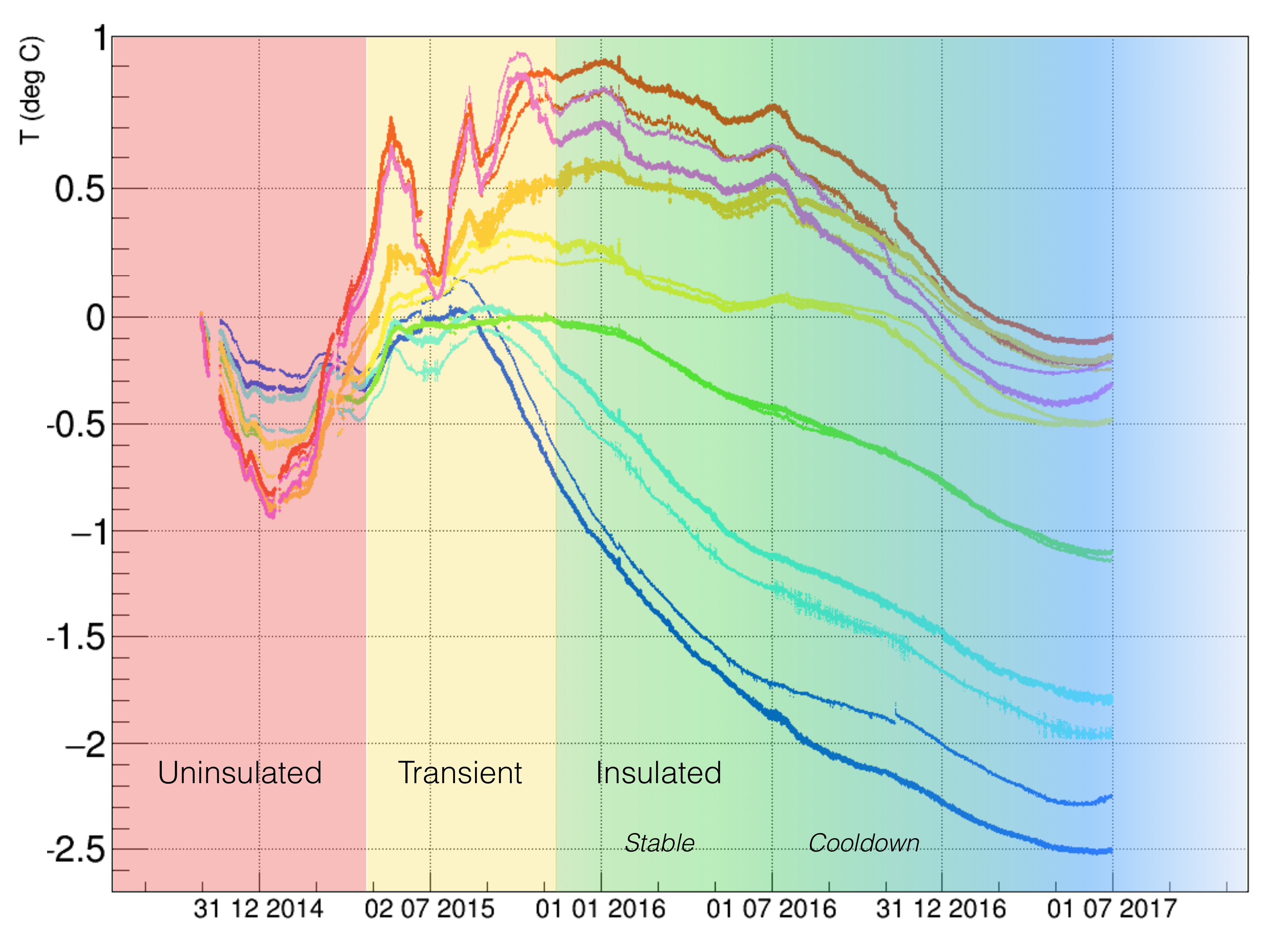}
\caption{Main periods in the October 29th, 2014 to October 21st, 2016 Phase I.a (OB) LTPS data-taking. Temperatures are offset-corrected to the same point at the start date, to then show relative drift. Color-coded curves (see online version of this work) capture the relative position of the LTPS sensors: progressively warmer colors (dark blue: $\approx$-67$^{\circ}$, cyan: $\approx$-50$^{\circ}$, green: $\approx$-26$^{\circ}$, yellow: $\approx$7$^{\circ}$, orange: $\approx$26$^{\circ}$, fucsia: $\approx$50$^{\circ}$ and red: $\approx$67$^{\circ}$) indicate the increasing SSS latitudes. Double lines are present because of the North-South sensor duplicity.}
\label{fig:periods}
\end{figure}

From the start of LTPS data taking in 2014 (see Figure~\ref{fig:periods}), until mid-January 2015, a large decrease in overall temperature, and also in top-bottom gradient was noted. Recorded data from the legacy thermometers show the minimum reached here ($\approx$2.2$^{\circ}$ between the extreme SSS instrumented latitudes $\pm$67$^{\circ}$) is probably the lowest ever reached since the start of Borexino data-taking in 2007. A rapid increase in gradient is then evident: while the bottom sees a slight increase of $\approx$0.15$^{\circ}$C, the top sees almost half a degree. This is considered the \textit{uninsulated} phase of the LTPS dataset. In May 2015, the TIS deployment started. This coincided with a rapid decrease in overall temperature and gradient -- although at first it was mostly motivated by environmental changes in the Hall, especially in the top temperatures decrease, since TIS surface coverage was still very minor. From July 2015, seasonal temperature rise in the Hall brought the temperatures up at the same time as the loop recirculator pump for the lower half of the water in the WT was shut down. This pump increased mixing in the lower WT, and with it heat exchange, disturbing stratification. Once residual currents died down, a large decay in temperatures in the bottom half of the LTPS was evident, including in the OB. This is considered the \textit{transient} phase, until autumn (around October) of 2015, when the TIS was deployed practically globally around the detector, and the bottom cooldown was well-established. The top-bottom gradient then started an almost uninterrupted hike until July of 2016, when it reached its all-time maximum of $\approx$5.2$^{\circ}$C. TIS coverage ensured that seasonal minima in environmental air temperatures affected much less the interior of the detector. The first half-year of this \textit{fully-insulated} period showed remarkable stability in all areas of the detector, except for the foreseen (and stabilizing, in the long run) bottom cooling down. August-September of 2016 brought with it a decrease in top temperatures, which caused a break in the approximately year-long increase of the gradient. AGSS operation was started in December 2016 with the aim of partially offsetting this trend, and locking the top temperature at a constant value that should never be surpassed by environmental conditions ($\approx$17.5$^{\circ}$C). The heater temperature setpoint was gradually raised in order to have the situation fully under control and avoid de-stabilizing top heating events that could perturb the background distribution in the IV. Meanwhile, the bottom temperatures are stabilizing to the thermal sink's equilibrium temperature of ~8$^{\circ}$C, and this situation is being translated to the inside of the SSS too: the \textit{fully-insulated} period will foreseeably be subdivided into \textit{cooling} and \textit{stably-stratified} subperiods. See Figure~\ref{fig:periods} for a graphical depiction.

For illustrative purposes, a na\"ive calculation can show the cooling time constant of an ideally-insulated detector (i.e., with adiabatic walls that let no external air influence seep in, and is therefore only constrained by the heat losses through the bottom). Although simplified, this calculation represents a worst-case scenario of detector-wide, irreversible cooling. Convection will not play an important role when the lowermost fluids are stratified -- therefore, a conduction-only scenario is a very good approximation to this case, where only faster heat transport along structures (through walls, legs...) may induce some small deviations by causing weak, localized convection. However, this should only cause localized "cold finger" structures, of no or little concern.

Taking the nominal C$_{H_2O}$=4186 J/(kg$\cdot$K) and C$_{scint}$=1723 J/(kg$\cdot$K), and considering we get a mass of 280 tonnes (IV) + 1040 tonnes (OV) = 1320 tonnes of scintillator, as well as 2100 tonnes of water in the WT, we can estimate the total detector's heat capacity as:

\begin{equation}
1.32 \cdot 10^6 kg \cdot 1723 \frac{J}{kg \cdot K} = 2.27 \cdot 10^9 J/K 
\end{equation}
\begin{equation}
2.1 \cdot 10^6 kg \cdot 4186 \frac{J}{kg \cdot K} = 8.8 \cdot 10^9 J/K
\end{equation}
\begin{equation}
C_{BX}^{total}=11.1 \cdot 10^9 J/K
\end{equation}

Using the lowermost Phase II.a (and -67$^{\circ}$ I.a) sensors, which show an approximately-linear temperature drop for short enough periods ($\approx$months), and extrapolating these trends to volumes at approximately the same temperature at each corresponding height, the worst-case heat loss through the bottom heat sink (since the data points were chosen at the beginning of the fully-insulated phase, when the inner fluids are still warm, and furthermore at the start of winter) can be estimated at $\approx<$250 W = 250 J/s $\approx$ 7.9$\cdot$10$^9$ J/year$\approx$0.3$^{\circ}$C/year. Although more realistic estimates will be shown later, this simple calculation using $\Sigma \Delta T\cdot C^{PC/H2O}_{BX}$ provides a useful upper limit.

The largest temperature gradient occurs in the bottom half of the detector, as evidenced by Phase II sensor data, changing from the approximately-constant 8$^{\circ}$C at ground level (kept stable by the aquifer located in the rock under Borexino, at $\approx$6$^{\circ}$C) to $\approx$14.5$^{\circ}$C around the equator ($\approx$9 m higher). The stratification is, generally, much less defined in the top half, although ever present.

\section{Initial conductive CFD modeling}
\label{sec:conductive}

A proper understanding of the thermal and fluid-dynamic environment stemming from foreseeable developments in the regions of interest inside Borexino was needed during and after the BTMMS deployment and operation. These studies are also necessary to ensure the highest internal thermal and fluid stability through AGSS operations. Furthermore, development of a robust numerical study would expand the LTPS' monitoring capability (point-like) to a detector-wide scale.

The commercial finite-volume solver ANSYS-Fluent (v.16.2 and 17.0) was employed for Computational Fluid Dynamics (CFD) simulation purposes in the CFDHub of the Politecnico di Milano's Computational Center. Initially, full-detector simulations employing only conduction, following the Q$^3$ guidelines\cite{riccardo}, will be reported. These obeyed the desire to (i) characterize the large-scale temperature differences in the detector configuration, with respect to a motionless "ideal" stratification, (ii) study long-term temperature trends in the full detector under different boundary conditions, (iii) study the boundary effects of the TIS and AGSS, and (iv) identify the importance of the conducive role that structures may have as thermal bridges through the fluid. Two- and three-dimensional models using reference LTPS data were employed -- even though a 2D simulation was adequate for most cases, the 3D case was an important verification for it, providing a way to study thermal transport along structures in a realistic geometry.

However, as is obvious, these conductive cases would not account for convection or fluid movement, imposing a clear limitation in simulating thermal trends away from stable stratifications or solid structures, as well as in understanding where backgrounds would be led by fluid-dynamical currents. 

\subsection{Numerical domain and meshes}

\begin{figure}[ht]
\centering\small\includegraphics[width=1\linewidth]{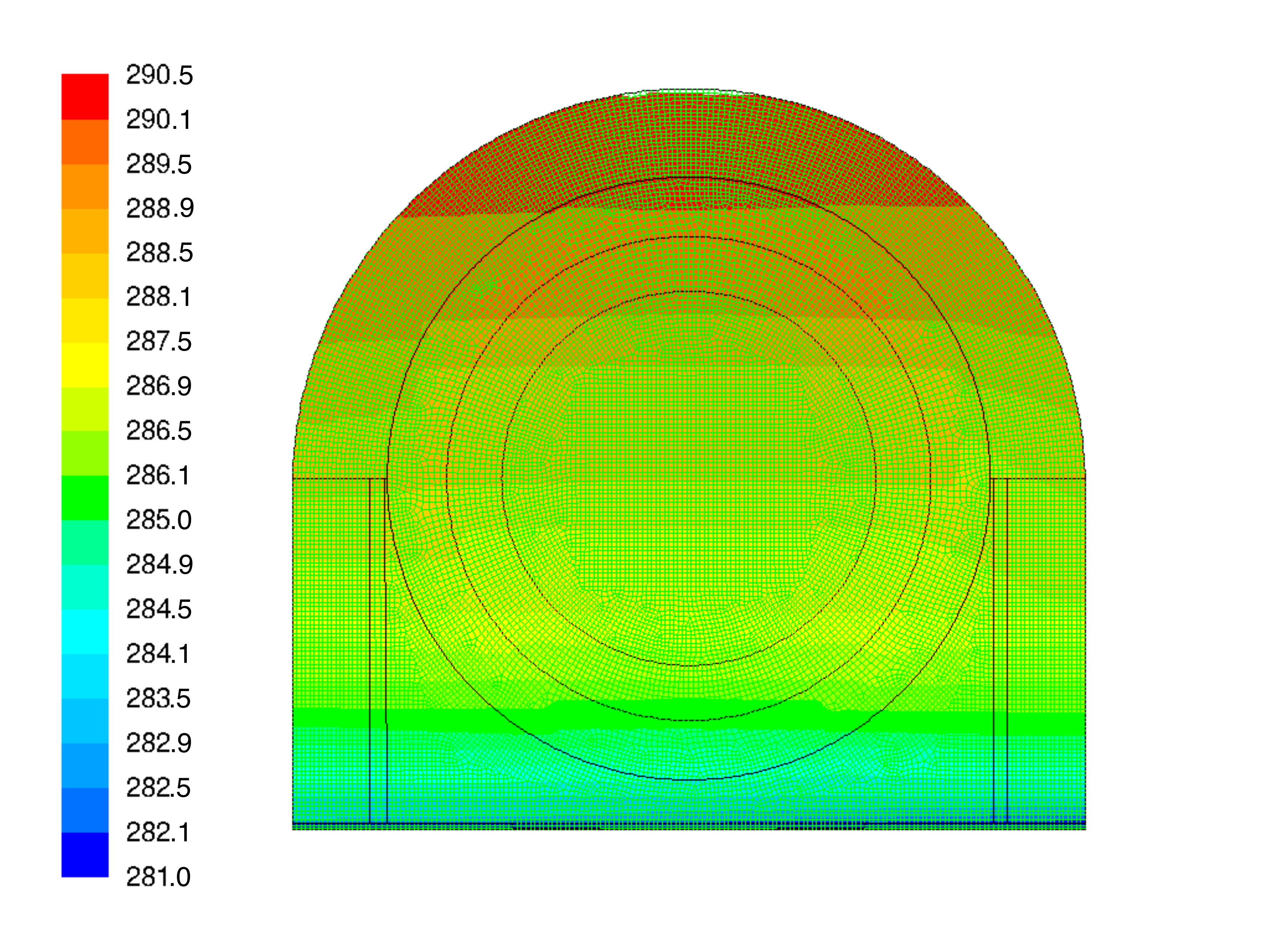}
\caption{Elements of the model used for the 2D conductive simulation of the temperature distribution of Borexino. Superimposed are the mesh and the isotherm temperature profile resulting from initialization conditions based on LTPS data from 2015/12/15, at the end of the "\textit{Transient}" period (in K). The Z coordinate origin is located at the base of the SSS legs, centered on the model's symmetry axis.}
\label{fig:mesh}
\end{figure}

The 2D model depicted in Figure~\ref{fig:mesh} follows a Cartesian coordinate system and includes the following structural elements: a Water Tank boundary (steel, 1 cm thick, 16.9 m high, 18 m OD), an SSS boundary (steel, 8 mm thick, 13.7 m OD), a North and South leg (steel, 14.3 mm thick, 32.4 cm OD, water-filled) supported on an equatorial platform (steel, 1 mm, see footnote\footnote{This unrealistic thickness was chosen because of the grilled nature of the platform: even though the actual thickness is $\approx$2 cm, the porosity is estimated at $\approx$90$\%$}), Inner and Outer vessels (nylon, 125 $\mu$m thick), and bottom plates (steel, 10 cm thick for the upper one on top of a smaller 4 cm thick lower one) supported on a concrete platform (14 cm maximum thickness). Water fills the WT interior and the SSS/vessels are filled with pseudocumene. An imposed temperature boundary condition mimicking the heat sink measured with the LTPS Phase II.b probes (8$^{\circ}$C) is set on the base of the model. WT wall boundary conditions vary depending on the scenario. Worst-case AGSS operation (in the sense of maximal undistributed heating) is simulated with a constant-temperature 2.1 m-long band which can be turned on or off around the proper height.

The mesh is based on 10 cm-side square cells on the bottom half of the water tank, transitioning to radial on the spherical dome and interior of the SSS. To avoid inaccuracies and instabilities in the region of interest in the center of the detector, a rectangular cell pattern was established again in this area, motivating irregular transition cells at around 3 m radius. In the two-dimensional model, the structural elements are just barriers with a given thermal resistance, but do not conduct along their length: this characteristic is only possible in the 3D model.

\subsection{Initial and boundary conditions}

The model was subdivided in a series of 15 different "domains", according to the height separations established by the positions of the LTPS' Phase I.a (OB), I.b (WT) and II.a (WT wall) probes. These domains were divided in \textit{North} or \textit{South} sides (and \textit{Center}, if applicable, for the volume inside the SSS). A linear interpolation was established between the known-temperature points the probes are located at. The discontinuity between domains is left to happen at roughly the position of the SSS to avoid unphysical temperature jumps happening at the most interesting regions (FV, WT periphery). The interpolation functions are:

\footnotesize
\begin{equation}
\begin{split}
T^{N/S}(x,y)=\frac{1}{A} \big[ (\Delta T(R_1) |x-|x_0|| + \Delta T(R_0) |x-|x_1||) y + \\
+ |x-|x_0|| (T_1y_1 - T_2y_0) + |x-|x_1|| (T_4y_1-T_3y_0) \big]
\end{split}
\label{eq:interp}
\end{equation}
\begin{equation}
\begin{split}
T^C(x,y)=\frac{||x|-R_0|}{A} \big[ (\Delta T^S(R_0)+\Delta T^N(R_0))y + \\
+ (T^S_4+T^N_1)y_1 - (T^S_3+T^N_2)y_0 \big]
\end{split}
\label{eq:interp2}
\end{equation}
\normalsize

where $A$ is the area between 4 interpolation points (where the temperatures $T_{1...4}$ are known from the LTPS sensors), $x_{0/1}$ and $y_{0/1}$ are the locations of each interpolation point, and $\Delta T$ is the temperature difference on each side of the interpolation square for the WT walls ($R_{0/1}$) or the SSS walls ($N/S$) at the appropriate distance from the model's vertical axis: $R_0$ refers to the Sphere's radius, $R_1$ to the WT's. These interpolation functions were properly generalized for a 3D case, in order to have a smoothly-varying temperature in all directions, through the addition of a $z$ dependence in order to account for a smooth radial function.

\subsection{Results}

\paragraph{2D conductive}

Conduction-dominated cooling was verified through the use of different boundary condition scenarios on the WT walls: ideally-insulated walls (adiabatic) and realistically-insulated walls (20 cm-thick TIS) with a constant, or time-varying temperature profile (whose $\pm$2.5$^{\circ}$C modulation could be weighted or unweighted with height, since it is observed the temperature of the warmer air on the top of Hall C oscillates with larger amplitude than the thermally more stable bottom, colder air). All external air boundary conditions are based on LTPS Phase III.c and legacy sensors. A control case with no insulation was also run for a seasonally-varying gradient boundary condition. All models were run for a year of simulated time, and use the initial thermal profile derived from the LTPS readout for December 15th, 2015.

\begin{figure*}[ht]
\centering\small\includegraphics[width=0.7\textwidth]{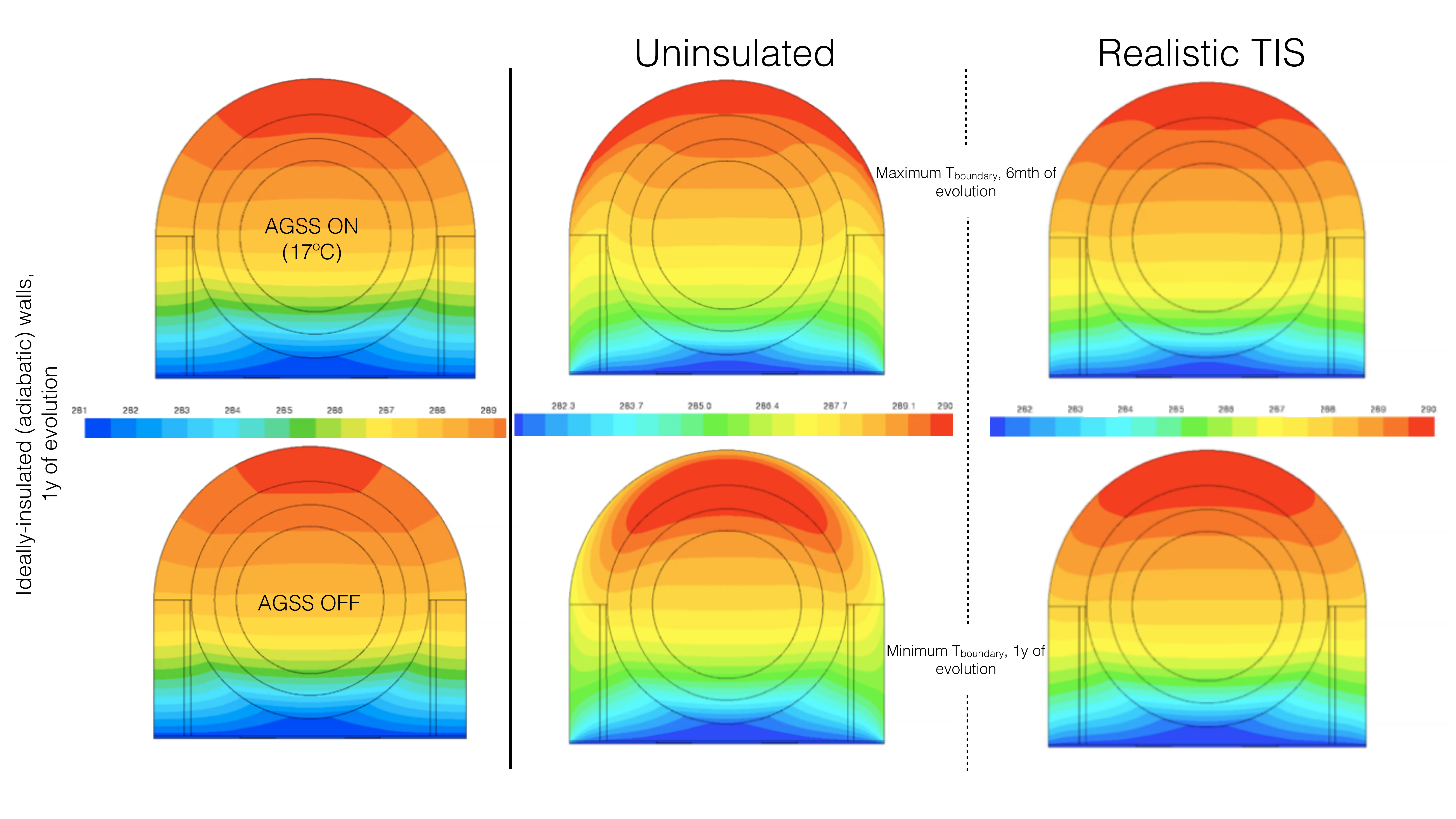}
\caption{Isotherms of conduction-dominated simulations. The left two figures show the stabilizing effect of the AGSS in the ideally-insulated (adiabatic) detector walls, slowing top cooling after a year of evolution. The four other figures show the difference between the height of warm (top row, 6 months of simulated time) and cold (bottom row, 1 year of simulated time) exterior boundary condition temperatures. In particular, the center two figures depict the uninsulated case with height-weighted $\pm$2.5$^{\circ}$C external oscillating boundary conditions, at their yearly maximum (top) and minimum (bottom). Identical conditions are imposed on the model on the right, which is however simulated with a TIS-like insulation layer on its exterior walls, making the seasonal changes in the periphery much milder. Scale spans 281 K to 290 K. Color version available online.}
\label{fig:evolution}
\end{figure*}

Similar behaviors could be seen in the center of the detector, where a "lenticular" convex feature (see Figure~\ref{fig:evolution}) grew on the cold, bottom isotherms, initially flat and horizontal, no matter what the thermal behavior was in areas closer to the boundary. This cold front progressed upwards until reaching the SSS, whereupon the isotherms regained horizontality again.

\begin{figure*}[ht]
\centering\small\includegraphics[width=0.6\textwidth]{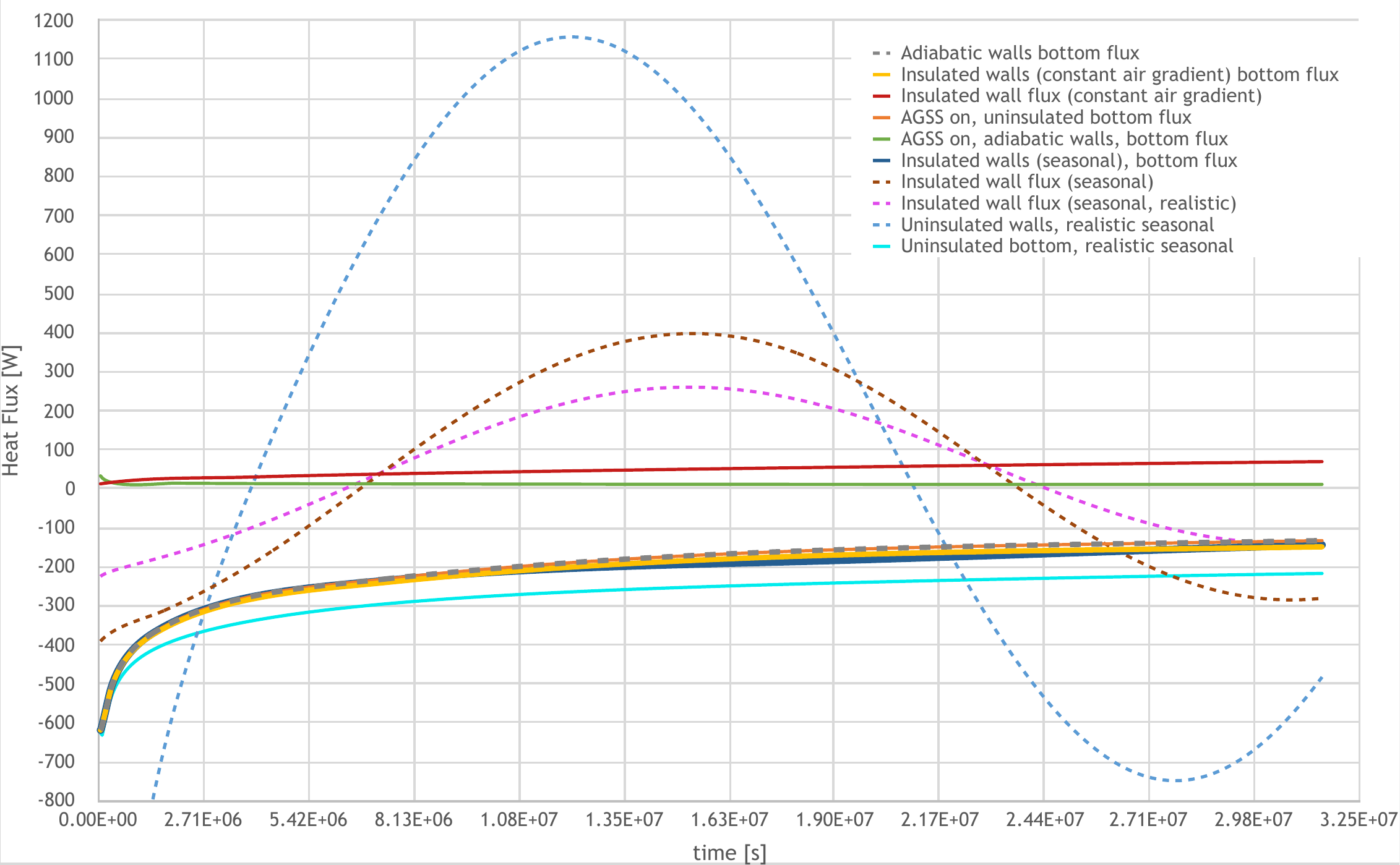}
\caption{Absolute heat fluxes through the 2D model's surfaces in different cases. Definite negative curves show heat flux through the bottom, including the perfectly-insulated/adiabatic case (discontinuous grey line), the uninsulated case (continuous light blue line) and the realistically-insulated case with constant (continuous yellow line) and seasonally-varying (continuous dark blue line) exterior temperatures. The bottom flux with adiabatic walls and activated AGSS overlaps with the one without AGSS operation at this scale (orange line under the previously indicated discontinuous grey one). Definite positive curves show instead the AGSS heat flux absorbed by the tank (continuous green line), as well as the heat exchange with a constant-temperature external environment (red). Finally, the oscillating dashed curves show heat flux through the walls in realistically-insulated cases (brown, seasonal change unweighted with height; pink, weighted), as well as the uninsulated control case (light blue, weighted seasonal change with height).  This figure's color version is available online.}
\label{fig:powerlosses}
\end{figure*}

\begin{figure}[ht]
\centering\small\includegraphics[width=1\linewidth]{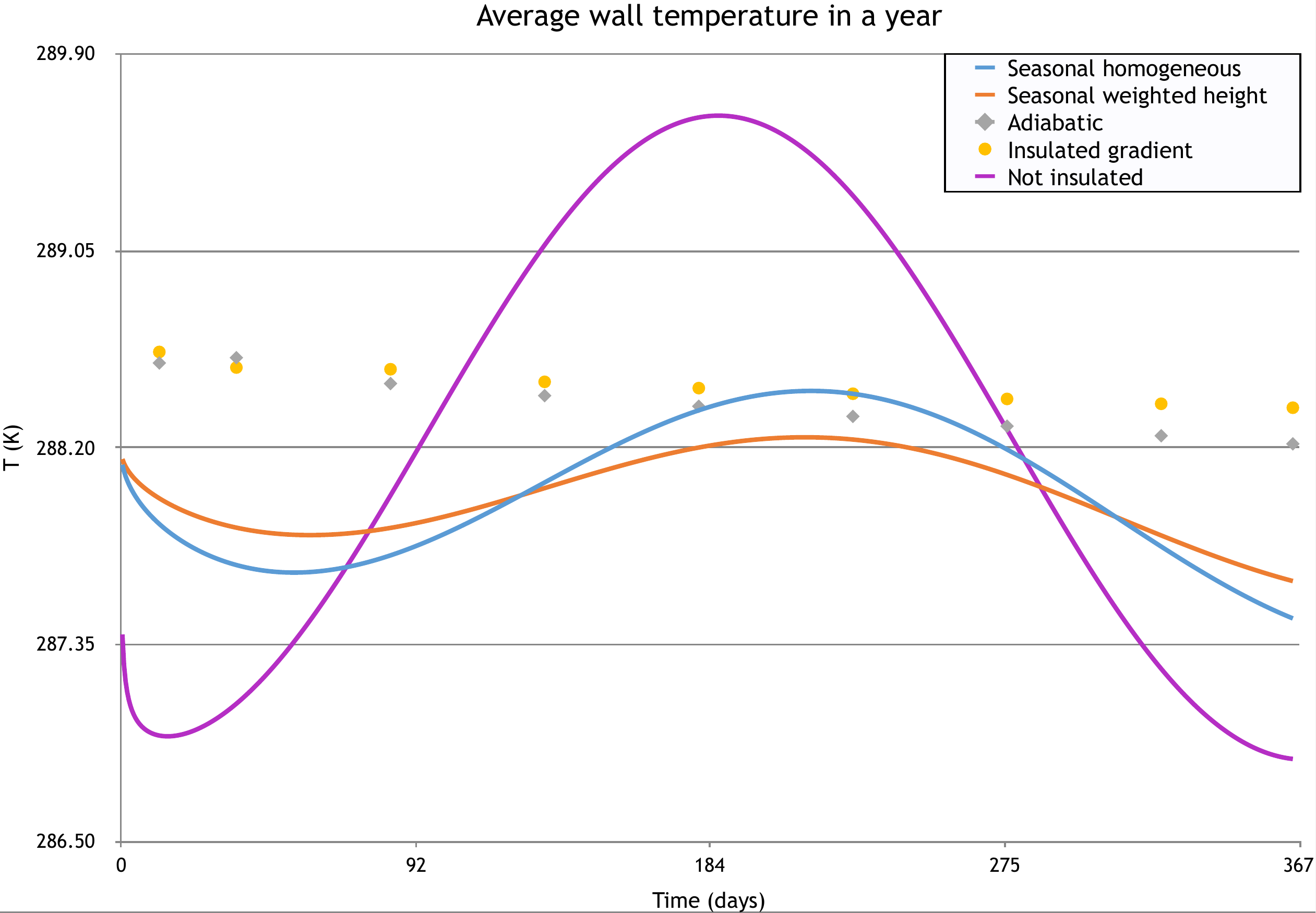}
\caption{Average wall temperatures highlighting the difference between the seasonal exterior boundary condition that changes equally at the bottom than at the top (blue) and the one weighted with height, changing maximally at the top and minimally at the bottom (orange). An uninsulated control case with weighted seasonal temperature change is also shown in purple. Dotted lines represent the scenarios with adiabatic walls not accepting any exterior heat influx or loss (grey points) and with realistically-insulated walls surrounded by a constant exterior air temperature (yellow points). This figure's color version is available online.}
\label{fig:powerlosses2}
\end{figure}

This provides a strong confirmation that, even in a worst-case situation where fluid cannot move to redistribute the heat transfer, the bottom cooling through the heat sink boundary condition retains a strongly local character, as opposed to a global effect in which the whole detector temperature drops strongly when being decoupled from the external environment by the TIS. Indeed, the only scenario where the cooling becomes global in the detector, is the one with completely adiabatic walls, as could be expected. Even in this case, the regional characteristics of the bottom cooling are retained, and only a relatively minor cooldown is seen on the top regions. As mentioned, inside the SSS, and especially in the FV, the temperature profile exhibits very minor differences between the adiabatic or realistically-insulated cases with different boundary conditions. In Figures~\ref{fig:powerlosses} and~\ref{fig:powerlosses2}, specific and integrated heat transfer time profiles are shown, identifying a maximum of $\approx$70-80 W equilibrium difference between the extreme cases of adiabatic and uninsulated walls, and the almost coincident bottom heat transfers in all insulated cases.

As mentioned in the previous section, an estimate of a maximum of $\approx$250 W of heat loss through the bottom was determined from the temperature drop in the lowermost Phase I probes during a short period of time -- which allowed for linearization of the trend -- in the December 2015-January 2016 interval, when the detector-wide TIS installation was just finished. This estimate is well in agreement with the equilibrium heat loss through the bottom derived from these conductive models, which (after the initial $\approx$month of model stabilization from the interpolated profile) show a maximum of $\approx$-300 W, plateauing out at less than 200W of lost power in the long run.

These results yield confidence for a conduction-dominated bottom cooling that should provide a maximum temperature drop of $\approx$-0.3$^{\circ}$C/year, as estimated in Section~\ref{sec:exp_results}, with a trend that slowly reduces this drop's magnitude. This cooling is furthermore restricted to the bottom reaches of the WT, and has a negligible effect on Borexino's mid-to-upper reaches.

Differences in the time profile for the temperature inside the SSS were small in all cases ($\ll \mathcal{O}$(0.1)$^{\circ}$C) and almost insignificant except for the uninsulated case. However, these conduction-only results solely provide upper limits on the thermal transport inside the Sphere, and should not be quantitatively interpreted, as convection-dominated processes will accelerate thermal transport there. Around the boundaries though, thermal behavior is much more conduction-dominated. There, temperature variations were much more prominent, where a clear "flapping" effect was seen on the isotherms (whereupon the isotherms' vertical displacement is much more pronounced close to the boundaries, and gets reduced away from them), especially on the uninsulated case, as was expected from the smoothing of outer boundary disturbances through the behavior of Poisson's equation away from those boundaries. Indeed, the interior isotherms were remarkably stable throughout seasonal variations in this conductive-only scenario, but water temperatures in the OD exhibited large oscillations that were dampened by $\approx$65$\%$ for the models with the TIS.

Stabilization of the upper gradient was also attempted by means of a simulated AGSS (two constant-temperature arcs on the location of the serpentines, simulating a maximal heating model). We imposed the approximate initial temperature at that height: 17$^{\circ}$C, as AGSS setpoint, and considered a worst-case cooling scenario with perfectly-insulated WT walls (adiabatic). Although the net effect in power loss through the bottom was almost negligible, highlighting the regionality of bottom cooling versus top heating, the homogenization of temperatures was less evident in the top, and kept a more ample high-temperature area around the top, effectively "holding" the top isotherms in place, although the actual heat input from the AGSS to the top water is extremely small ($\approx$0.2 W/m$^2$).

Another relevant insight is that the system, at the time of initialization, is remarkably stable conductive-wise. No major changes occur in the temperature distribution apart from the bottom cooling, a tightening of the isotherms in the lower part of the detector, and boundary effects due to the heat exchange with the environmental air, in the cases where such exchange is allowed. SSS interior temperatures remain relatively unchanged. Of course, given the symmetry of the boundary conditions, the initial asymmetry between the North and South sides disappears quickly -- nevertheless, this apparently minor feature will be shown to have great importance in the convective case. In any case, the robustness of this conductive model's results already hint at the near-stability condition the detector exhibits at the start of the fully-insulated period.

\paragraph{3D conductive}

The three-dimensional model allowed for an in-depth study of heat conduction along boundaries, which was not possible in the 2D case, along with a confirmation of the validity of the 2D model's trends. Revolution surfaces were created from the 2D mesh boundaries along the model's vertical axis, except for the legs, which were individually modeled (only 14, since it was expected they would not provide an important contribution to heat transfer). The pit under Borexino was also modeled, but its presence should not be relevant to the simulation's outcome, since only the ceiling Phase II.b sensors were considered for this setup, imposing the bottom heat sink at 8$^{\circ}$C on the WT's floor. The exterior boundary condition was kept adiabatic, although as mentioned heat conduction along the skin of the wall is possible. Two scenarios were run: with the AGSS on at 20$^{\circ}$C (the maximum foreseeable range of operations) and with the AGSS off.

Worst-case (maximal heating with homogeneous fixed-temperature band along the WT's 6th ring) AGSS operation was confirmed to remain restricted to the heated band and very narrow border areas around it, not posing any possibility of the heat creeping downward through the tank's wall (plausible in principle, given the metal's higher thermal conductivity) in a problematic fashion. A conductive-only scenario such as this should also provide a further layer of conservatism to this result, given that convection in the water should keep the heat even more localized in reality. Therefore, the AGSS system is reliably shown to offer safe operation within its intended design objectives.

Other structures that could offer a heat exchange path (legs and equatorial platform) were seen to exchange minimal heat with its surroundings, not increasing or decreasing significantly the heat conduction of the materials that surround them. This is again a worst-case scenario, since convection would only limit the heat transmission though these structures when fluids move around them. It is shown no additional heat transfer and, with it, convection, will occur because of the legs or platform, therefore making it safe to ignore them as heat-conductive elements in convective models -- thereby reducing their complexity and allowing for increased efficiency. 

Overall thermal profiles were identical to the equivalent two-dimensional cases, which therefore are considered fully reliable for conductive-only scenarios.

\begin{figure}[ht]
\centering\small\includegraphics[width=1\linewidth]{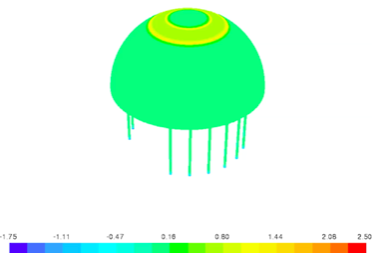}
\caption{Near-equilibrium condition of the 17$^{\circ}$C setpoint of the AGSS after a year of operation and perfect (adiabatic) insulation. Although heat transfer was greater at the beginning of the operation, because of initial transients, non-zero heat transfer never extended beyond the limits visible in the image. Color version available online.}
\label{fig:AGSS}
\end{figure}

\section{Convective CFD Methodology}
\label{sec:methodology}

In order to move beyond the lower limits obtained from the conductive simulations and validate the convective CFD approach \cite{riccardo} in reproducing the main characteristics and phenomena inside Borexino (closed system, Newtonian water-like fluids, prevalent natural convection, vertical temperature difference ($\approx$10$^{\circ}$C) and thermal stratification), several benchmarking cases have been modeled and their results compared with the available experimental data.

\subsection{Design parameters}

The choice of dimensions and $\Delta T$ for the following benchmark models needed to be motivated to at least lie close to, or ideally overlap, Borexino's regime of interest. The determination of the Rayleigh number for Borexino offers the simplest, most rigorous way of relating seemingly dissimilar geometries to the detector case. The definition of the Rayleigh number is very dependent on the model geometry, and in non-standard ones (such as Borexino's spheric geometry with distributed, gradual temperature differences) may be somewhat arbitrary if not keeping a close watch on the phenomenon under study. The Rayleigh number ($Ra$) is a dimensionless parameter defined, in general, as:

\begin{equation}
Ra = \frac{\beta [K^{-1}] \Delta T [K] g [m/s^2] L^3 [m^3]}{\nu ^2 [m^4/s^2]} Pr
\label{eq:Rayleigh}
\end{equation}

where $\beta$ is the thermal expansion coefficient of the fluid, $\Delta T$ is the temperature difference in the characteristic lengthscale of the system, $g$ is the gravitational acceleration acting on the system, $L$ is the characteristic lengthscale for natural convection in the system, $\nu$ is the kinematic viscosity of the fluid and $Pr$ is the Prandtl number, which is itself defined as the quotient between the momentum diffusivity and the thermal diffusivity. In practice, the Prandtl number is only dependent on the fluid's nature and state. The quotient multiplying $Pr$ is referred to as the \textit{Grashof number} $Gr$, which is a measure between the buoyancy and viscosity forces on a fluid.

As can be inferred from this definition, the Rayleigh number is most dependent on the characteristic lengthscale $L$ for convection in the considered system. $Ra$ is therefore a way of relating buoyancy-driven fluid flows coming from different fluids, conditions and system geometries -- and therefore, contains information about the convective/conductive dynamics of a fluid flow, irrespective of the fluid. This is in contrast to the Grashof number, which depends upon the fluid under consideration.

If we consider the typical overall gradient in Borexino's IV to be $\approx$5$^{\circ}$C, over the 8.5 m between the top and bottom poles of the vessel, we get $\approx$1.7 m/$^{\circ}$C: that is, $\approx$17 cm separating each 0.1$^{\circ}$C isotherm. Considering this is our $L$, $Ra \approx \mathcal{O}(10^7-10^8)$ (with $Pr$=7.78 for PC, $\beta _{PC}^{10C} \approx$10$^{-3}$ K$^{-1}$, and $\nu_{PC}^{10C} \approx$7$\cdot$10$^{-7}$ m$^2$/s). We consider the $\mathcal{O}(0.1)^{\circ}$C temperature differences routinely happening in short timescales in Borexino, which may be causing the internal stirring concerning us. If the overall gradient were very large, the isotherms would be very close together, and a given $\Delta T$ seeping in from the outside would show up at a smaller lengthscale than if the overall gradient was smaller, and the isotherms were farther apart from each other -- in which case the isotherm displacement to match the boundary condition would occur over larger lengthscales.

We are of course assuming a linearly-stratified fluid, which is not the real case in Borexino (which exhibits a laxer stratification on the top than on the bottom). Therefore, we should keep in mind the order-of-magnitude Rayleigh number estimate above would be approximately 1-2 order(s) of magnitude larger, locally, on the top, and smaller on the bottom. Consequently, we can estimate \textit{Borexino's Rayleigh range} as $Ra$ $\epsilon [\mathcal{O}(10^6),\mathcal{O}(10^9)]$.

\subsection{Governing Equations}
\label{governingequations}

The finite volume commercial solver \textit{ANSYS-Fluent} is used for modeling the flow field via mass, momentum and energy conservation equations for incompressible Newtonian fluids with constant viscosity and density. Governing equations of mass, momentum and energy are numerically treated using the Direct Numerical Simulation (\textit{DNS}) approach for laminar flows, as reported in Eq. \ref{mass}, \ref{eq:momentum} and \ref{energy}, respectively.

\begin{equation} \label{mass}
\frac{\partial\rho}{\partial t}+\nabla\cdot\left(\rho\mathbf{u}\right)=0
\end{equation}

\begin{equation} \label{eq:momentum}
\frac{\partial{\rho\mathbf{u}}}{\partial t}+\nabla\cdot(\rho\mathbf{uu})=-\nabla p+\nabla\cdot\bar{\bar{\tau}}+\bar{\rho}{\mathbf{g}}
\end{equation}

\begin{equation} \label{energy}
\frac{\partial \rho E}{\partial t}+\nabla\cdot(\mathbf{u}(\rho E+p))=\nabla\cdot(k\nabla T+(\bar{\bar{\tau}}\cdot\mathbf{u}))
\end{equation}

where \textit{\textbf{u}}, \textit{p} and \textit{E} represent the velocity vector, static pressure and energy, respectively, $\bar{\rho}\mathbf{g}$ the gravitational body force, $\kappa$ the thermal conductivity and $\bar{\bar{\tau}}$ the stress tensor \citep{Malalasekera2007}. For the natural-convection flows, the Boussinesq model used here models fluid density as a function of temperature: it treats density $\rho$ as a constant value in all solved equations, except for the buoyancy term in the momentum equation, where it is approximated as:

\begin{equation} \label{boussinesq}
\bar{\rho}\mathbf{g}=(\rho-\rho_{0})\mathbf{g}\approxeq -\rho_{0}\beta(T-T_{0})\mathbf{g}
\end{equation}

where $\beta=-\frac{1}{\rho}(\frac{\partial \rho}{\partial T})_p$ is the thermal expansion coefficient, $\rho_{0}$ and $T_{0}$ the reference values for density and temperature, respectively.

\subsection{Geometrical modeling}

The computational mesh used to discretize the domain is a structured Cartesian or unstructured polygonal/polyhedral grid, depending on the geometry. Specific refinements near the walls are applied to take into account the viscous and thermal boundary layer. The mesh size ($\Delta x$) is defined for each modeled geometry and it is based on a preliminary mesh sensitivity analysis. This permitted to quantify the influence of different grid sizes selecting the largest acceptable size, with computational grids ranging from $\mathcal{O}(10^{4})$ to $\mathcal{O}(10^{5})$ elements.

\subsection{Numerical modeling}
\label{sec:Numerical modeling}

All domains modeled here are closed and boundary conditions are imposed, including friction and shear stress on adjacent fluids, for the fluid dynamics, and considering adiabatic or fixed external temperature for heat transfer conditions. 

The solver used to perform the transient simulations is based on the coupling pressure-velocity PISO algorithm \citep{Malalasekera2007}. It is able to guarantee the convergence at each time step, through inner loops, using a restrictive Courant-Friedrichs-Lewy (\textit{CFL}) condition ($CFL\leq1$) to maintain the necessary accuracy \citep{Malalasekera2007}. The time-step size $\Delta t$ is defined considering the physical $\Delta T_{p}$ and numerical Fourier stability analysis $\Delta T_{Fo}$ natural convection constrains:

\begin{equation} \label{physical}
\Delta t_{p} = \frac{\tau_{0}}{4} \approx \frac{L}{4\sqrt{\beta g L \Delta T}}
\end{equation}

\begin{equation} \label{numerical}
\Delta t_{Fo} = \frac{Fo(\Delta x)^{2}}{\alpha}
\end{equation}

with $\tau_{0}$ the time constant, $L$ the characteristic length, $Fo$ the Fourier number (limited to $Fo=0.1$), $\Delta x$ the grid size and $\alpha=\frac{k}{\rho c_{p}}$ the numerical diffusivity. Based on fluids properties, operational conditions and geometrical characteristics the maximum time-step size has been defined for each simulation presented below.

The same discretization schemes have been used for all simulations, as reported in Table \ref{tab:numerical schemes}.

\begin{table}[h]
\begin{centering}
\begin{tabular}{|c|c|}
\hline
\textbf{Term} & \textbf{Scheme}\tabularnewline
\hline
\hline
Transient & First Order Implicit\tabularnewline
\hline
Gradient & Least Squares Cell Based\tabularnewline
\hline
Pressure & Body Force Weighted\tabularnewline
\hline
Momentum, Energy & Third Order MUSCL\tabularnewline
\hline
\end{tabular}
\par\end{centering}
\caption{Numerical Schemes \label{tab:numerical schemes}}
\end{table}

\section{Convective benchmarking validation}
\label{sec:benchmarking}

\subsection{Phenomenological benchmarks}
\label{sec:phenobench}

The level of currents that represent a concern inside Borexino's IV is derived from the background's half-life ($\tau_{1/2}^{^{210}Po} \approx$ 138 days) and the IV's dimensions (4.25 m nominal radius, or $\approx$1 m from the vessel where the $^{210}$Pb progenitor sits and provides an "inexhaustible" $^{210}$Po source for our purposes). This means that currents under $\mathcal{O}$(10$^{-7}$) m/s would be so slow that more than half of the detached polonium will decay away in the trip, even under directly radial motion. Therefore, the level of admissible numerically-induced systematic uncertainties should not exceed this magnitude and ideally be $\leq \mathcal{O}$(10$^{-8}$) m/s.

Simple scenarios involving a cylindrical 2D geometry were implemented to characterize the basic phenomena at work in a well-studied scenario with a regular square mesh that ideally avoids the creation of preferential directions. This rectangular mesh grid features an average cell size of $\approx$3 cm (11 cm$^2$). Initial conditions for all cases were set as a linearly-stratified temperature gradient of [10,18]$^{\circ}$C according to:

\begin{equation}
T(h)=T_2 + (T_1-T_2) \frac{h-h_0}{H}
\end{equation}

where H is the cylinder's height (13.7 m, to keep it within Borexino's dimensions, along with its width of 11.2 m) and $T_1$ ($T_2$) is the top (bottom) temperature.

A scenario where no motion would be expected established the level of irreducible numerical noise for the model at $\lesssim$3.5$\cdot$10$^{-7}$ m/s, with an undefined, random pattern across the model.

Sudden temperature variations on the bottom (increasing $T$) and/or top (decreasing $T$) surfaces, keeping the walls with an adiabatic boundary condition, showed regional effects circumscribed to those areas, which extended only until the height of the corresponding isotherm was reached. Threshold for convection triggering recirculation cell formation was $\Delta T >$0.1$^{\circ}$C. Equivalent dynamics were found for both top and bottom. Largest achieved currents were $\mathcal{O}$(dm/s). This result proves the inherently safe principle of operation for the AGSS based on heat application on Borexino's top dome.

In contrast, the application of non-adiabatic boundary conditions to all boundary surfaces, including the walls, prompted the generation of a global convective mode spanning the whole cylinder geometry. It showed varied characteristics depending on the $\Delta T$, but always organized around robust currents of rising/falling fluid along the walls, and weaker recirculation currents along the central axis, with the addition of meandering horizontal currents or recirculation cells for large $\Delta T$. Uniform, height-weighted or delayed (through the addition of varying thicknesses of insulation on the model's boundaries) $\Delta T$s were employed to characterize the different behaviors -- nevertheless, an important final conclusion is that there is no "allowable" threshold on the amount of temperature difference that would not induce a global circulation pattern, if $\Delta T/\Delta t$ is small. Some currents enter the realm of the resolution limit for the model $\mathcal{O}$(10$^{-9}$-10$^{-8}$ m/s), and would not be relevant in the case of Borexino, but the organized convective structure remains in place.

\paragraph{Literature models and conditions} Reproduction of experimental results from selected literature examples were identified as the proper benchmarking strategy. The selected benchmarks \cite{Yin, Garg, Goldstein} are characterized by two-dimensional geometry, with cylindrical section (or spherical, but it is identical in a two-dimensional circular section assuming azimuthal symmetry) and the presence of a concentric annulus fully contained inside, with different ratios between the inner and outer radii, as shown in Table~\ref{table:benchmark_conditions}. The inner annulus' outer surface is the one at high temperature, while the exterior annulus' inner surface is the one at low temperature, for all cases presented here. The analysis focuses on the fluid behavior in the space between both, and it is initially considered at a constant, volume-weighted, mean temperature defined by:

\begin{equation}
T_m = \frac{(r_{av}^3-r_i^3)T_i + (r_o^3-r_{av}^3) T_o}{r_o^3-r_i^3}
\end{equation}

where $r_{av}$ is the average radius $(r_o+r_i)/2$ and $r_i$ ($T_i$), $r_o$ ($T_o$) are the inner and outer radii (temperatures), respectively\cite{Yin}.

The numerical geometry for the benchmarking cases is a 2D model of the real geometry, with a structured mesh including a local refinement close to the walls. In all geometries the minimum cell size $(\Delta x)_{min}$ ranges from $2\cdotp10^{-4}$ to $5\cdotp10^{-5}$ m and the number of cells from $2.3\cdotp10^{4}$ to $3.68\cdotp10^{5}$ for the coarse and fine meshes, respectively. Numerical methods and algorithms are those described in Section \ref{sec:Numerical modeling}, with a time-step size $\Delta t$ ranging from $2$ to $9$ s for the fine and coarse meshes, respectively.

For the first cases, taken from~\cite{Yin}, the Grashof numbers given in the reference were converted to the dimensionless Rayleigh number; this was taken as reference to calculate the inner/outer surface temperatures, as well as the fluid's parameters according to Equation~\ref{eq:Rayleigh}. This operation involved an amount of (informed) arbitrariness, since the literature did not indicate the absolute temperatures they worked with. For that reason, ranges around the 10-20$^{\circ}$C relevant for Borexino were chosen when possible. The fluid employed was water, since the reference parameters are much better constrained than for PC/benzene at the small $\Delta T$s involved. Sometimes, owing to the low $Ra$ used, the temperature difference for a high-viscosity fluid such as water would be too small ($<\mathcal{O}(10^{-3})^{\circ}$C), and air was chosen instead. Dimensions were kept as in the reference. 

\begin{table*}[t]
\centering
\begin{tabular}{p{1.5cm} p{1cm} c c c c}
\multicolumn{2}{c}{\textbf{Medium}} & \multicolumn{4}{c}{\textbf{Characteristics}} \\
\cmidrule(r){1-2}
\cmidrule(r){3-6}
\textbf{Ref} & \textbf{Model} & \textbf{Ra} & \textbf{$r_o/r_i$ } & \textbf{$\Delta T$} (K) & \textbf{Features of interest} \\
\cline{1-6}
Air\cite{Yin} & Air & 5880 & 1.78  & 0.64 & Chimney \\
Air\cite{Yin} & Air & 5880 & 1.4 & 2.33 & Chimney + upper cells\\
Water\cite{Garg} & Water & 90000 & 2 & 0.24 & Isotherm and circulation pattern \\
Air\cite{Garg} & Water & 250000 & 2 & 0.11 & Isotherm and circulation pattern (kidney-shape) \\
Air\cite{Yin} & Air & 739200 & 2.17 & 0.02 & Kidney-shaped cells + elongated cells \\
Water\cite{Garg} & Water & 10$^6$ & 2 & 0.09 & Square-kidney pattern + upper and lower vortices \\
Water\cite{Yin} & Water & 1.5$\cdot$10$^6$ & 2.17 & 0.06 & Circulation pattern + vortices\\
Water\cite{Goldstein} & Air & 2.51$\cdot$10$^6$ & 2.6 & 9.26 & Chimney, vortex, flat isotherms \\
Water\cite{Yin} & Water & 7.1$\cdot$10$^6$ & 1.78 & 0.33 & Distortion of steady flow into unsteady vorticity \\
Water\cite{Yin} & Water & 10$^7$ & 1.78 & 2.3 & Small circulation features in medium-scale\\
 & & & & & pattern at change of regime: double vortex, \\
 & & & & & shear structure, chimney unsteadiness. \\
Water\cite{Yin} & Water & 21.6$\cdot$10$^6$ & 2.17 & 0.19 & Upper cell + detachment of outward current \\
\end{tabular}
\caption{Summary of the operating conditions for the different annuli benchmarking simulations. The ratio between external and internal radii is $r_o/r_i$, and $\Delta T$ is the temperature difference imposed between the inner (hotter) and outer (colder) annuli.}
\label{table:benchmark_conditions}
\end{table*}

\paragraph{Results}

A summary of the achieved results, based on the main relevant features in each of the benchmark cases, is included in Table~\ref{table:benchmark_summary}.

\begin{table*}[t]
\centering
\begin{tabular}{p{1.5cm} p{1cm} c c c c}
\multicolumn{2}{c}{\textbf{Medium}} &  & \multicolumn{3}{c}{\textbf{Features}} \\
\cmidrule(r){1-2}
\cmidrule(r){4-6}
\textbf{Ref} & \textbf{Sim} & \textbf{Ra} & \textbf{Small} & \textbf{Medium} & \textbf{Large} \\
\cline{1-6}
Air\cite{Yin} & Air & 5880 & - & \textbf{Chimney} & \textbf{Crescent} \\
Air\cite{Yin} & Air & 5880 & Upper vortices & \textit{Flow direction} & \textbf{Crescent} \\
Water\cite{Garg} & Water & 90000 & - & \textit{Cell center} & \textbf{Crescent, isotherms} \\
Air\cite{Garg} & Water & 250000 & - & \textit{Cell center} & \textbf{Kidney, isotherms} \\
Air\cite{Yin} & Air & 739200 &  \textit{Vortex structure} &  \textit{Cell center} & \textbf{Kidney, isotherms} \\
Water\cite{Garg} & Water & 10$^6$ & \textit{Vortices} & \textit{Streamlines} & \textbf{Isotherms, fluid flow} \\
Water\cite{Yin} & Water & 1.5$\cdot$10$^6$ & Stronger vortices & \textit{Upper flow} & \textbf{Fluid flow} \\
Water\cite{Goldstein} & Air & 2.51$\cdot$10$^6$ & \textbf{Heat transfer} & \textit{Isotherms} & \textbf{Nusselt} \\
 & & & \textbf{Nusselt} & &  \\
Water\cite{Yin} & Water & 7.1$\cdot$10$^6$ & Stronger vortices & \textit{Transition threshold} & \textbf{Fluid flow} \\
Water\cite{Yin} & Water & 10$^7$ & \textit{Double vortex} & \textbf{L-shape} & \textbf{Fluid flow}  \\
 & & & \textit{and shear} & \textbf{detached features} & \textbf{and structure} \\
Water\cite{Yin} & Water & 21.6$\cdot$10$^6$ & \textit{Vortex position} & \textbf{Vortex} & \textbf{Flow} \\
 & & &  & \textbf{detached features} & \textbf{stagnant region} \\
\end{tabular}
\caption{Summary of CFD literature benchmarking results, showing well-reproduced features (\textbf{boldface}), features present with small deviations from literature (\textit{itallics}) and features which are absent or present with large deviations (normal font). Illustrations of some of these features can be found in Figures~\ref{fig:Yin_21600000_217}, \ref{fig:Garg1M} and~\ref{fig:Yin_1492800_217}}
\label{table:benchmark_summary}
\end{table*}

We can confront the average Nusselt number $\overline{Nu}$ from the inner cylinder's exchanged power (59.654 W). The Rayleigh number is the dimensionless number to compare natural convection between different cases, while the Nusselt number compares the heat exchange. From the reference, we know that:

\begin{equation}
\overline{Nu}_{conv} = \frac{\overline{h}_i D_i}{\kappa}
\label{eq:nusselt}
\end{equation}

where $\overline{h}_i$ is the local heat transfer coefficient in the inner cylinder, $D_i$ is its diameter and $\kappa$ is the thermal conductivity. We also know that $Nu = \overline{\kappa}_{eq} \cdot Nu_{cond}$, and since $Nu_{cond} = 2 / \ln(D_o/D_i)$=2.09, we can calculate the $Nu$ from the $\overline{\kappa}_{eq}$=7.88 provided in the reference's Table 1 (for our $Ra$=2.51$\cdot$10$^6$). Indeed, the total average $Nu$ is then $Nu$=7.88$\cdot$2.09=16.47.

We shall compare this dimensionless number to the one obtained through the data from the simulation, because the Nusselt number offers a way to find a correspondence between very different cases, from the point of view of geometry and fluids, with regard to heat exchange. It is noted the heat transfer coefficient cannot be the same as in the reference case because we use water instead of air for reference fluid, among other model considerations. However, the fact that the model is not precisely equal to that in the paper increases confidence in its validity, since we expect behavior to be largely equivalent between models with the same Nusselt numbers, in spite of changes in the nature of the fluid employed.

In any case, from the numerical point of view, we have a different $\overline{\kappa}_{eq}$, and we want to obtain the $\overline{Nu}$ shown in Equation~\ref{eq:nusselt}, so we need $\overline{h}_i$, defined as:

\begin{equation}
\overline{h}_i = \frac{Q}{\pi D_i Z (T_i-T_o)}
\end{equation}

where $Q$ is the total power exchanged and $Z$ is the depth of the cylinder (20.8 cm)\cite{Goldstein}. The $\Delta T$ is 9.26$^{\circ}$C. With this data, we have $\overline{h}_i$=276. This yields a $\overline{Nu}$=16.431 (with a thermal conductivity $\kappa$=0.6 W/(m$\cdot$K)), which is in very good agreement with the $Nu$ found with the reference data.

\begin{figure}[tb]
\centering 
\includegraphics[width=1\columnwidth]{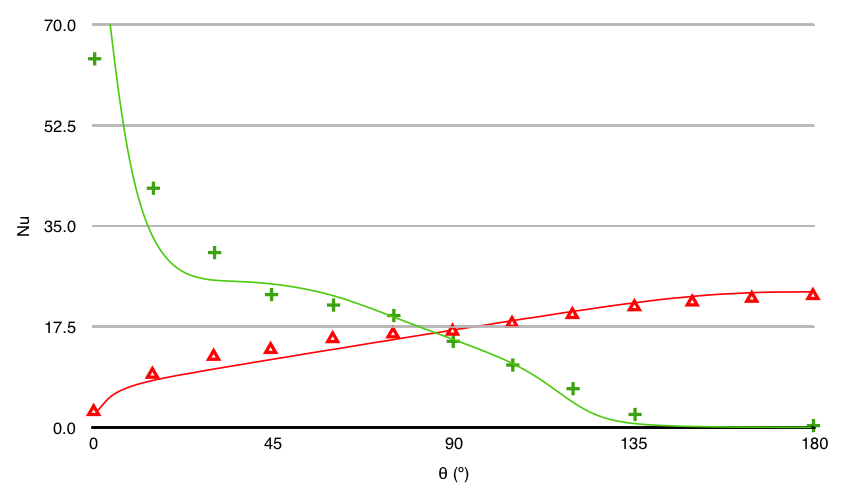} 
\caption{Comparison of the dimensionless local Nusselt numbers, between this paper's CFD results (solid lines) and the reference's\cite{Goldstein} (data points), for the inner (red, approximately flat trend line under Nu=20) and outer (green, decaying line close to zero for $\theta >$135$^{\circ}$) annulus surfaces. Color version available online.}
\label{fig:Goldstein_heattransfer} 
\end{figure}

We can also compare the $h$ for different points along the inner/outer cylinder's surface to establish a comparison with Figure 8 in the reference, which shows the local heat transfer coefficient versus the angular position considered, from the total surface heat flux (W/m$^2$) shown in Figure~\ref{fig:Goldstein_heattransfer}. Values were converted to Nusselt number in order to establish a one-to-one comparison irrespective of differences in the fluid.

From the values reported, we can readily see the curves follow similar trends, as can be appreciated in Figure~\ref{fig:Goldstein_heattransfer}.

In conclusion, the benchmarking showed good reproducibility of the thermal environment (when available to compare in the references) as well as the large- and medium-scale features present in each of the cases. Furthermore, while some small-scale features were not well-reproduced in some of the lowest Rayleigh number cases (and thus further from Borexino's regime), the general fluid flow pattern was faithfully reproduced in practically all regions of all cases.

\begin{figure}[tb]
\centering 
\includegraphics[width=1\columnwidth]{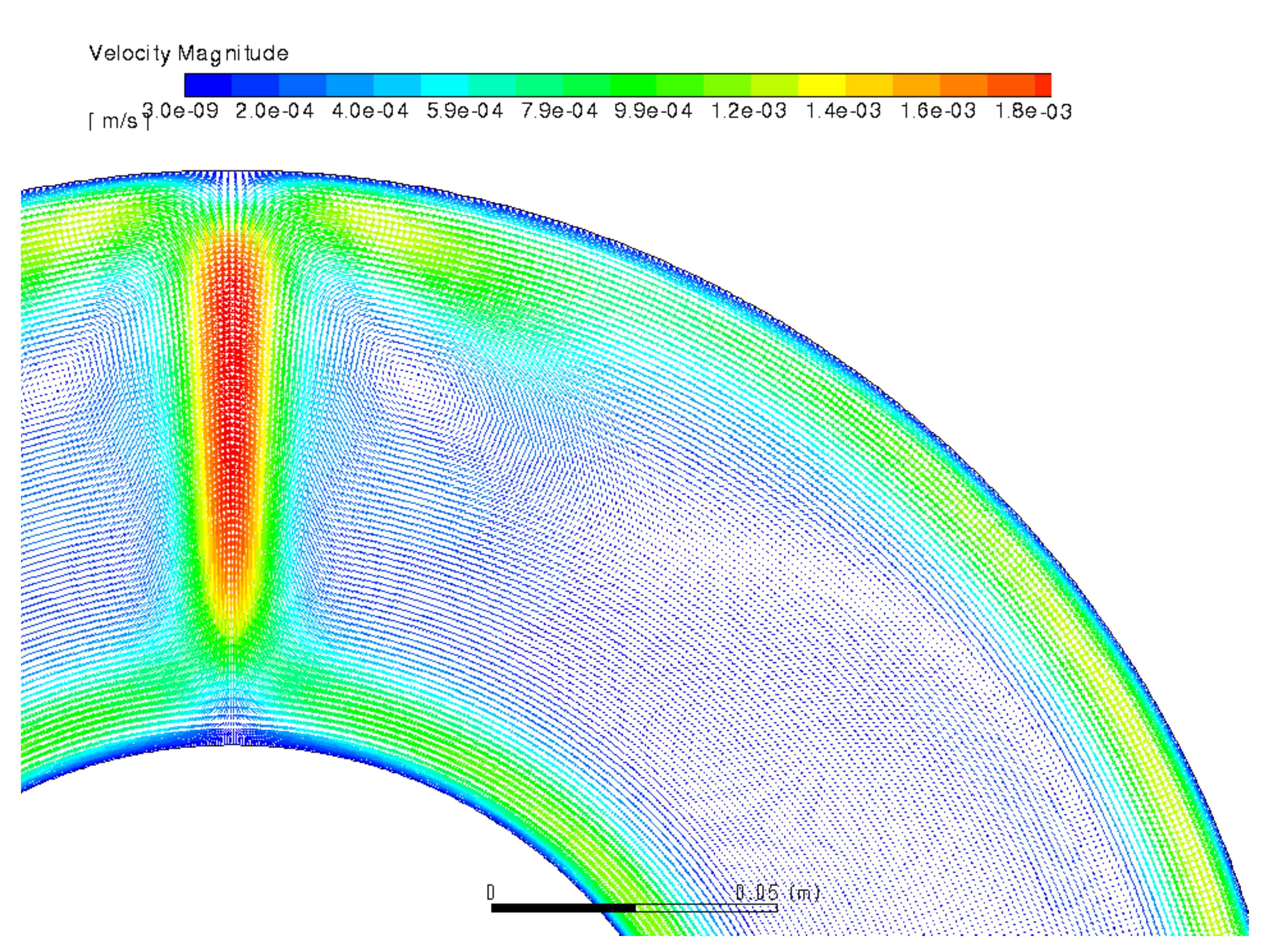} 
\caption{$Ra$=21600000; $D_o/D_i$=2.17 (water) showing the unsteady flow recirculation pattern featuring an upper weak recirculation vortex and the relatively stagnant inner fluid condition, accompanied by the high speed flows on the spheres and enlarged "chimney" structure. Units in m/s. Color version available online.}
\label{fig:Yin_21600000_217} 
\end{figure}

\begin{figure}[tb]
\centering 
\includegraphics[width=1\columnwidth]{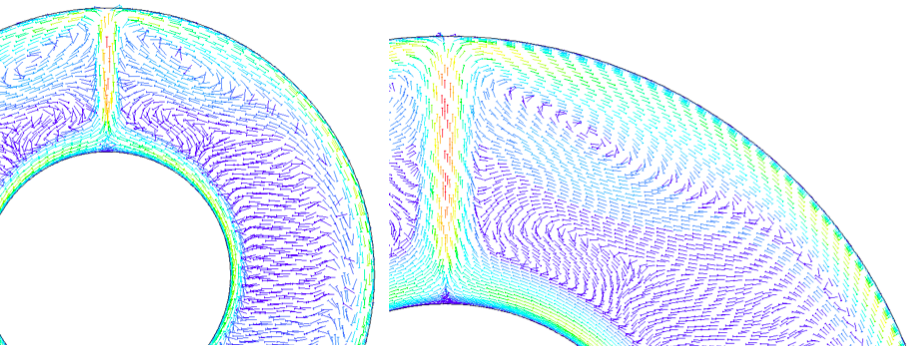} 
\caption{$Ra$=1492800; $D_o/D_i$=2.17 (water) showing the basic steady flow recirculation pattern in general (left) and in detail (right), featuring a stronger vorticity than in the reference case, but a good agreement with the general circulation pattern, especially at the angle between the upper circulation layer on the inner sphere and the "chimney" feature. Color version available online.}
\label{fig:Yin_1492800_217} 
\end{figure}

\begin{figure}[tb]
\centering 
\includegraphics[width=1\columnwidth]{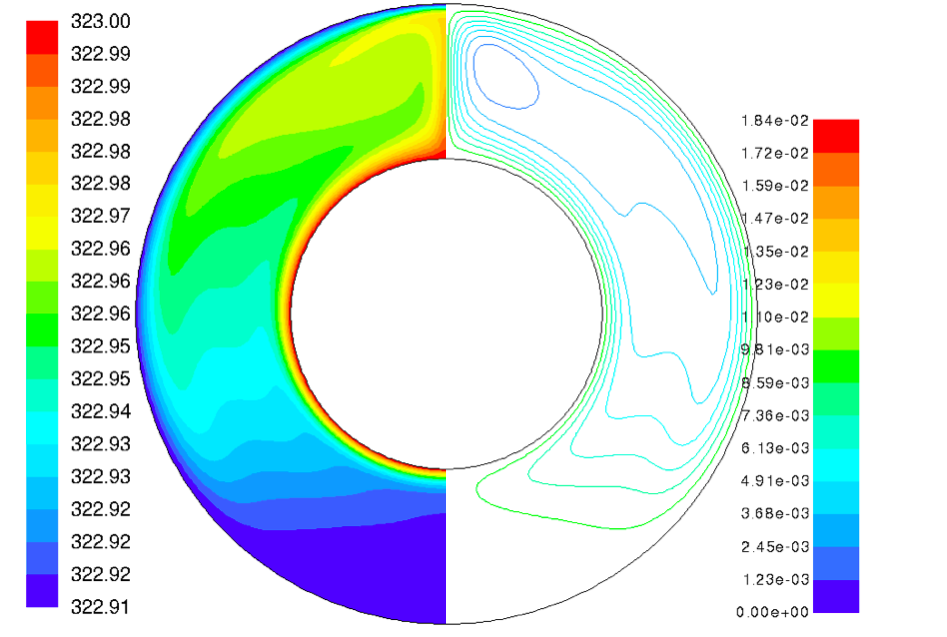} 
\caption{$Ra$=1000000; $D_o/D_i$=2.6 (water) showing the isotherms agreeing well with the general features seen in the interferogram for this case in \cite{Garg} (left, in K), as well as the circulation pattern (right, in kg/s). The upper vortex is clearly seen, as is the "squared kidney" shape of the main recirculation pattern. The center of the main pattern is shifted upward with respect to the reference, and the lower weak vortex at 180$^{\circ}$ is not visible, although a correct flow separation occurs and the main circulation pattern is a good fit to the reference. Color version available online.}
\label{fig:Garg1M} 
\end{figure}

\subsection{Borexino thermal benchmark} 

It is advisable to benchmark not only the general reproducibility of results just described, but also the model behavior in our particular cases of interest. Although we have no way of directly measuring fluid-dynamic effects inside the SSS, apart from the limited inference obtained from the background movement analyses, we do have a good thermal transport probing system: the Phase I LTPS sensors. We can employ the temperatures registered on the outside (\textit{i.e.} in the water around the SSS, through the Phase I.b probes) and study their transmission toward the inside of the Sphere. The most interesting feature here is that the simulated transmitted temperatures for the inside of the SSS can be confronted with the internal recorded temperatures (\textit{i.e.} in the buffer just inside the SSS, through the Phase I.a probes), and therefore establish the level of fidelity on thermal transport that the CFD strategy can offer.

The \textit{WaterRing} geometry is modeled as a 0.5 m-thick water volume around a PC-filled SSS segmented by spherical, rigid unsupported vessels. The water volume around the Sphere is a circular ring with its top/bottom areas truncated at the height of the SSS top/bottom, to avoid complications with interpolating the temperatures at higher/lower regions than the approximate measuring heights of the temperature probes. A separate initial condition was used for the water and SSS' interior, using the following linear interpolation:

\begin{equation}
T_{N/S}(t,y) = \frac{1}{h_1^i-h_0^i} \big( T_{N/S}^{i+1}(t) (y-h_0^i) + T_{N/S}^i(t) (h_1^i-y) \big)
\label{eq:boundary_t}
\end{equation}

where $h^i_{0/1}$ is the interpolation domain's upper(lower) height limit, which obviously coincides with the lower(upper) limit for the next $i+1$ (previous $i-1$) domain; $y$ is the vertical coordinate in any of the model's internal points, and $T_{N/S}^i$ are the recorded (North or South side) temperatures used as anchors on the corners of the square domains, taken from the recorded, time-varying LTPS Phase I(.a or .b) dataset. Care was exercised in order to keep the vertical coordinate of the LTPS probes the same as the domains' vertical limits, even though the horizontal position of the anchor point may vary slightly ($\mathcal{O}(10cm)$) to ensure smooth physical interpolation within the internal region of interest. 

The fact that this model is closed and enables fluid movement requires careful handling of the simulating conditions, in particular to the iterative timing, mesh geometry and iteration divergence probability.
 
To ensure an appropriate numerical modeling the mesh for the external part of the WaterRing benchmark has been meshed using a radial/Cartesian grid, with local refinement close to the boundaries representing the internal buffers. The internal part, representing the IV, has been meshed using both Cartesian and polygonal meshes, in order to check whose results show best accuracy and mesh-independence. Such analysis showed a dependence for the Cartesian grid, with preferential direction of the fluid inside the IV, and the total independence with the use of the polygonal mesh, which was therefore chosen. Based on such analysis and the guidelines carried out from the benchmarking cases reported in Section~\ref{sec:Numerical modeling} and \ref{sec:phenobench} the typical cell size $(\Delta x)$ is around $3\cdotp10^{-2}$ m, the number of cells $1.2\cdotp10^{5}$ and the time-step size $\Delta t$ equal to $9$s. This enabled for the best compromise between computational efficiency and expediency, and appropriate low numerical noise levels.
 
A custom-made software tool took care of checking, at each iteration, at which point the simulated time was, comparing it with the listed times in the recorded data from the LTPS probes. Once this simulated time reached or exceeded a given time limit (set at 1800 s, since that is the standard time delay between data acquisitions by the LTPS sensors), the appropriate recorded temperatures were updated as imposed boundary conditions on the water ring's surface. Provisions were implemented to ensure good data would always be available: in case of dropouts in DAQ, the imposed temperature would be kept at the last available time. This can cause a slight upset once new data is available, but the need to select relatively small periods for computation efficiency meant the dropout periods were short and few -- furthermore, these jumps were verified not to cause large enough deviations on the boundary conditions to motivate divergences or unphysical effects on the numerical solver. Conditions inside the SSS were left free to evolve when $t\neq0$, including on its boundary. This smoothed out the ring-to-SSS differences in initial conditions after a few numerical iterations.

This model's benchmarking power was realized by placing "tallying spots" in the nominal positions where the Phase I.a sensors would be. Simulated temperatures on these points, 1 m away from the imposed boundary condition, could be checked against the historical recorded data, to verify good thermal transport behavior across this instrumented region, representative of the whole Inner Detector.

\begin{figure}[tb]
\centering 
\includegraphics[width=1\columnwidth]{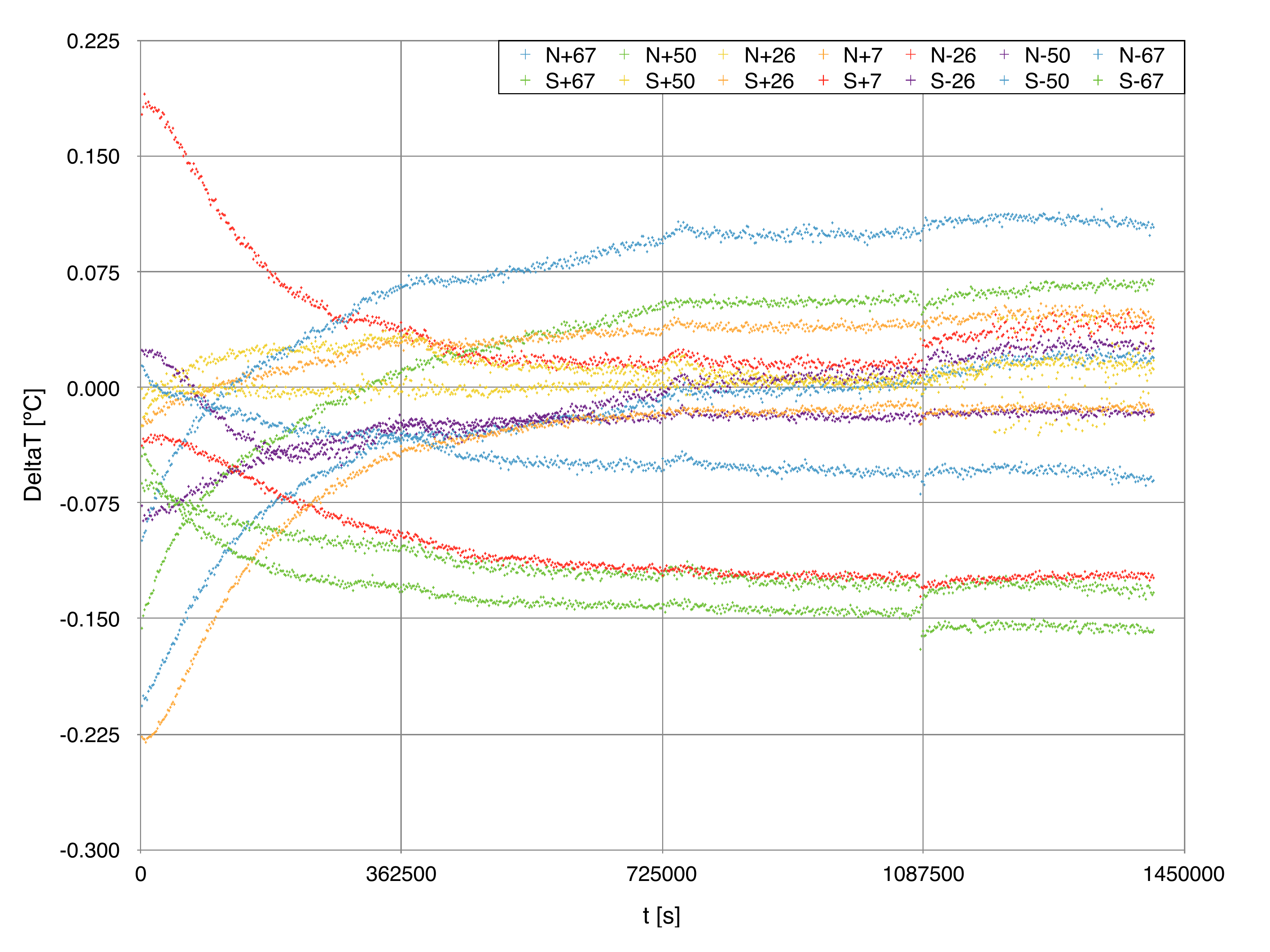} 
\caption{Residuals showing the difference between real recorded temperatures and simulated ones for the same locations during the same time interval (insulated period), for each of the 14 LTPS Phase I.a Outer Buffer temperature probes, tagged for clarity with their respective N/S side and latitude. It can be seen that, apart from initial settling transients between boundary and initial bulk conditions, they are stable and kept to less than 0.15$^{\circ}$C, showing the accuracy of the thermal transport simulation. Color version available online.}
\label{fig:residuals_WR_ins} 
\end{figure}

\paragraph{Results} Figure~\ref{fig:residuals_WR_ins} shows the residuals ("true" historical temperature minus simulated temperature at the same time and position) for the 14 positions of the LTPS Phase I.a probes. Good agreement can be seen ($\lesssim$0.225$^{\circ}$C), with a smooth exponential trend toward stable equilibrium errors no bigger than $\approx$0.15$^{\circ}$C. Discretization and temperature interpolation account for this level of errors. The overall trend shows a remarkable agreement between recorded and simulated data, as shown in Figure~\ref{fig:residuals_WR_ins}. Even more remarkably, the simulation shows a situation whereupon the initial temperature profile that sent temperatures to slightly incorrect values, due to the interpolation strategy followed, is corrected by the time evolution profile and allows the thermal trend to follow the recorded data profile within ~2 days of simulated time. It should be noted the 1-1.3 m/day thermal inertia measured in the conductive simulations accounts for at least part of this delay.

\begin{figure}[tb]
\centering 
\includegraphics[width=1\columnwidth]{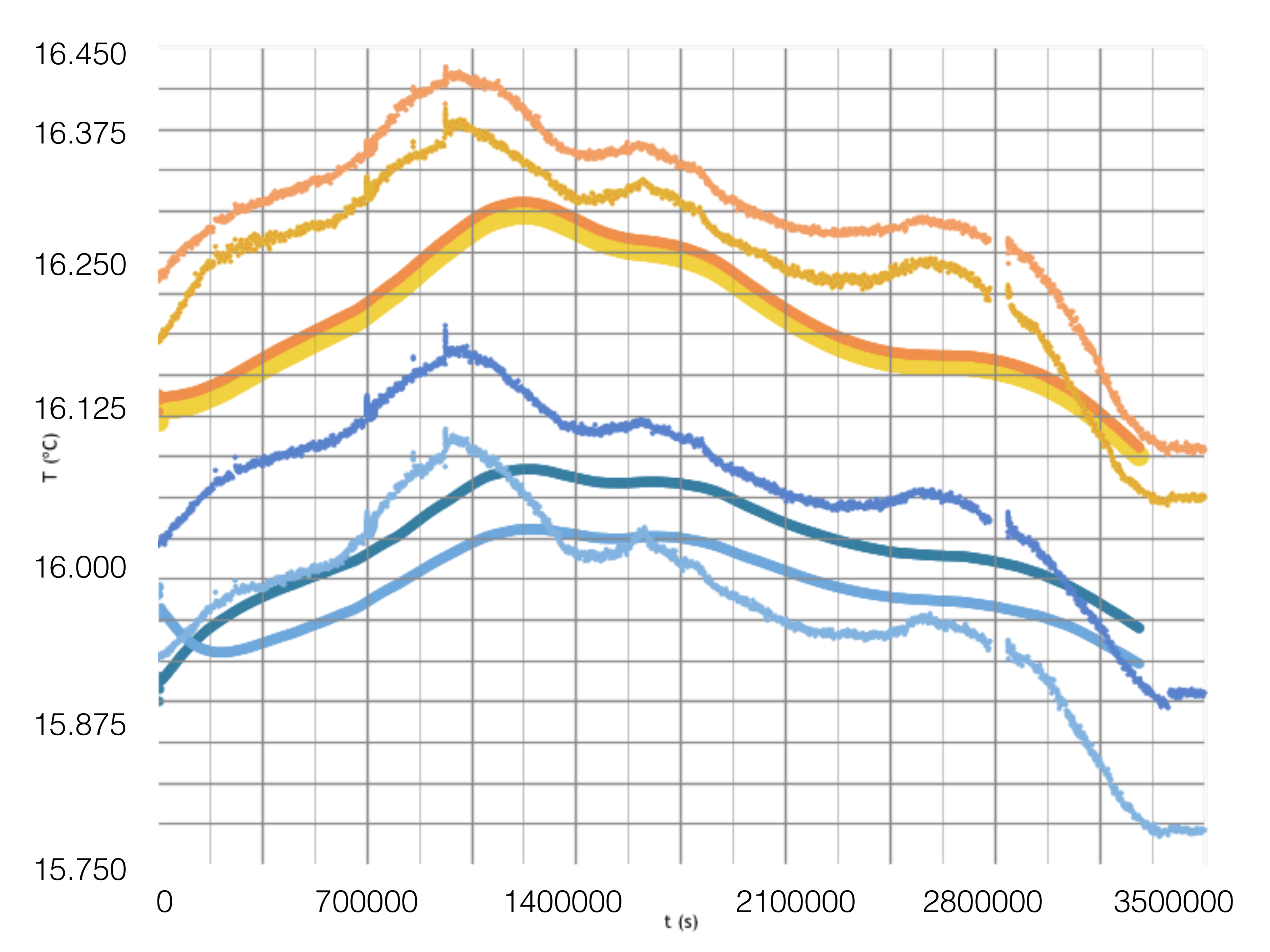} 
\caption{Detail of the real (jagged curves) and simulated (smoother curves) temperature profiles for the "Transient" time period considered and the two topmost Phase I.a LTPS sensors (67$^{\circ}$, or top two ragged/smooth curves; and 50$^{\circ}$, or bottom two ragged/smooth curves). A phase shift of $\approx$1-2 days is visible, apart from the obvious slight offset (maximal for these four sensors) in the abscissa axis. Color version available online.}
\label{fig:temps_WR_trans_detail} 
\end{figure}

A temporal phase shift is evident when focusing on the sharpest available features in the recorded temperatures and their simulated counterpart, as depicted, for example, in Figure~\ref{fig:temps_WR_trans_detail}. This shift is constant and the features (if the phase shift is manually corrected) are seen to line up almost perfectly, albeit with a certain, small decrease in slope change. The cause for this effect is still under investigation, but is considered not to negatively impact the overall reliability of the thermal transport benchmarking power of this model, given it represents a small and constant shift, and is moreover shown not to cause broadening of the features in the temperature trends.

The \textit{Water Ring} models are seen to provide a powerful benchmark for thermal transport, quite faithfully ($< \pm$0.2$^{\circ}$C, and much better in some cases) replicating the temperature evolution in the OB's LTPS Phase I.a probes positions when the boundary condition represented by the Phase I.b sensors is imposed $\approx$1 m away, in a different medium (water) and having to pass through the SSS structural element. Therefore, at least as far as thermal transport capabilities, the implemented \textit{ANSYS-Fluent} models are a useful tool to understand, replicate and foresee the thermal environment in the detector. Further, it is reasonable to believe that this extrapolation will hold, for the same geometry (and possibly for similar ones), at other points in the model. 

It is not, however, a benchmark for \textit{fluid} transport: there is no "ground truth" data from Borexino, since no "tracer" is available -- other than the same backgrounds we are trying to study through this research.

\section{Fluid dynamics in the ROI (SSS$\rightarrow$IV)}
\label{sec:SSS_IV}

The WaterRing benchmarking model was adapted to include just the Inner Detector, neglecting the water around it, and imposing the Phase I.a buffer probes readings on the SSS boundary, assuming a constant temperature from the real position of the LTPS sensors to the SSS surface, at the same vertical position. An increase of node density (with several cycles of mesh adaptation to improve performance and reliability) and a finer timestep of 4.5 simulated seconds per iteration was thus possible. CFD-derived temperatures were recorded at several points on the Inner Vessel in order to impose them as boundary conditions later on. Effectively, the benchmarking power of the \textit{Water Ring} models was employed to ensure the temperatures at the vessel boundary, where no probe is available, would be accurate. These measures should allow for minimization of systematic model-dependent errors, which could not be shown to be uncorrelated enough with the \textit{Water Ring} mesh geometries.

\subsection{IV-only}

\paragraph{Setup} The IV is modeled in 2D as a perfect circle of nominal Inner Vessel radius of 4.25 m. No attempt at modeling the actual vessel shape is made, although the deviations are small enough ($\leq \pm$20 cm, or $\approx$0.2$\%$) for this to be a good approximation. The polygonal mesh approach is used, with a cell size of $\approx$5 cm$^2$ ($\mathcal{O}(10^5)$ cells). No internal structures or localized mesh tightening is employed away from the model's boundary. Setting of initial conditions is performed picking the simulated temperatures a few centimeters outside the vessel simulated through the previously-described models. This small distance away from the vessel is chosen so as to avoid boundary layer effects that may locally shift the isotherms in a way that would falsify the most realistic temperature mapping in the bulk of the AV. As such, these temperatures were also used as input for a time-evolution script, in order to impose time-varying boundary conditions on the model's outer wall.

\paragraph{Boundary conditions} A perfectly-stratified model was first run in order to characterize the level of unphysical currents induced by the numerical iterative process, yielding a background level of $\approx \mathcal{O}(10^{-5})$ m/s. It is noted part of these currents, despite having a physical origin (especially in the horizontal direction), are mesh-enhanced. The absence of boundary currents along the vessel is notable, in sharp contrast with the model with time-changing realistic temperatures imposed as boundary conditions. The model's intrinsic numerical noise level, if the mesh can be regularized, can be much higher ($\mathcal{O}(10^{-8}-10^{-9})$ m/s), although the circular geometry of this model prevents a mesh that is completely regular all over the geometry. A three-dimensional model may be able to sidestep mesh-dependent, geometrical instabilities seen to develop in alternative checkerboard-patterned grids, but the large computational time required left this potential for numerical noise reduction as a future perspective.

Five time-dependent models were then run, chosen from the Phase I.b dataset in order to characterize the model in the most diverse array of thermal profiles as possible. This time span comprises the end of the "Transient" period and four regimes during the "Insulated" period, including the phase with the warm, stable top, and the later phase with a relatively-rapidly cooling top. The simulated time periods were between 2 and 8 weeks long.

\begin{figure}[tb]
\centering 
\includegraphics[width=1\columnwidth]{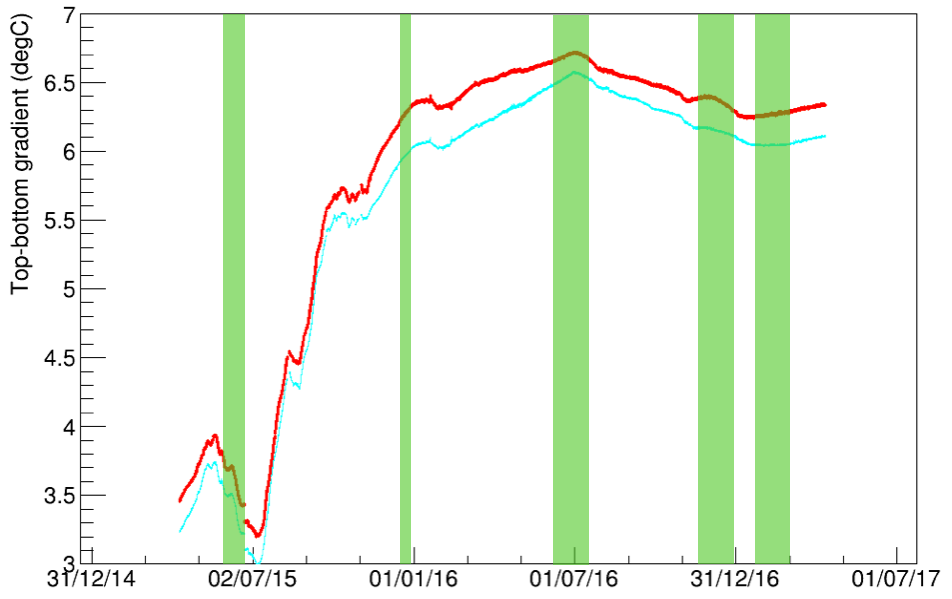} 
\caption{Time periods used in the IV simulations, superimposed on a plot of the (North or South, slim clear or thick darker line, respectively) vertical temperature gradient between the topmost and bottommost Phase I.b probes in the water around the SSS.}
\label{fig:wr_periods} 
\end{figure}

\subsection{Results} 

Horizontal currents are the dominant feature in all of the convective models considered with Borexino geometries and data-based temperature fields. Indeed, as expectable from the Rayleigh number calculation in Section~\ref{sec:methodology}, the large buoyancy potential energy gap between the top and bottom, separating the stably-stratified fluid layers, precludes bulk, organized motion in the IV. The spherical geometry, in contrast to the cylindrical symmetry from the cases in Section~\ref{sec:phenobench}, will have a role on this -- but asymmetries between both sides are the main driving force behind this preponderance of horizontal currents in place of vertical ones in realistic Borexino scenarios. Indeed, these asymmetries favor elongated recirculation cells that transport fluid from one side of the Sphere to the other one, while leaving the stratification in place, when they preferentially break organized vertical movement into minimal-energy transport along slightly-inclined isotherms.

\begin{figure}[tb]
\centering 
\includegraphics[width=1\columnwidth]{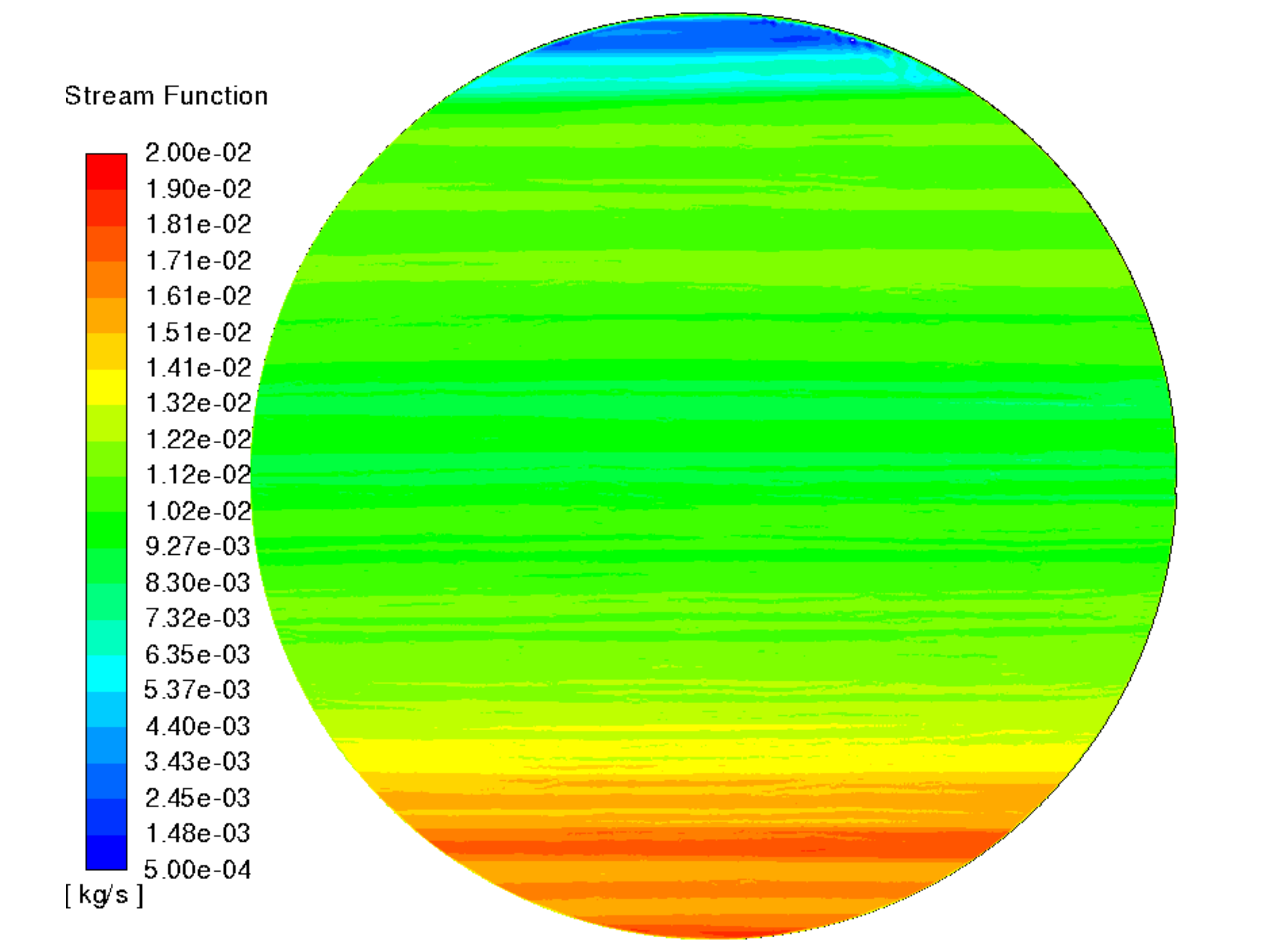} 
\caption{Typical iso-stream-function instantaneous map (in kg/s) for an IV-only model with time-varying boundary conditions based on realistic temperatures, showing fluid carrying capacity of the flow, and therefore indicating where the areas of maximum bulk transport are located. Color version available online.}
\label{fig:stft} 
\end{figure}

This is better observed through iso-stream-function views, which show the discretized local fluid-carrying capacity of the flow (see Figure~\ref{fig:stft}). They may allow for a better identification of the general fluid behavior, rather than studying the current velocities directly. Current velocities may skew the attention toward small, high-speed regions (which can be very localized and have no global importance, or even be numerically-induced by mesh irregularities). While these regions may propel fluid at the highest local velocities -- and with it, background radioisotopes -- , they need not account for most of the background injection into the FV -- indeed, often they are observed not to be correlated. On the other hand, current velocities may be useful for determining organized global movement (i.e. in case there was a vertical global "chimney" trend, as in the cylindrical benchmarks). However, we see no evidence of such movement in these cases. Individual tracking of virtual massless particles "attached to the fluid" through the \textit{pathline} ANSYS-Fluent \textit{Particle Track} utility provides confirmation of these observations.

\begin{figure}[tb]
\centering 
\includegraphics[width=1\columnwidth]{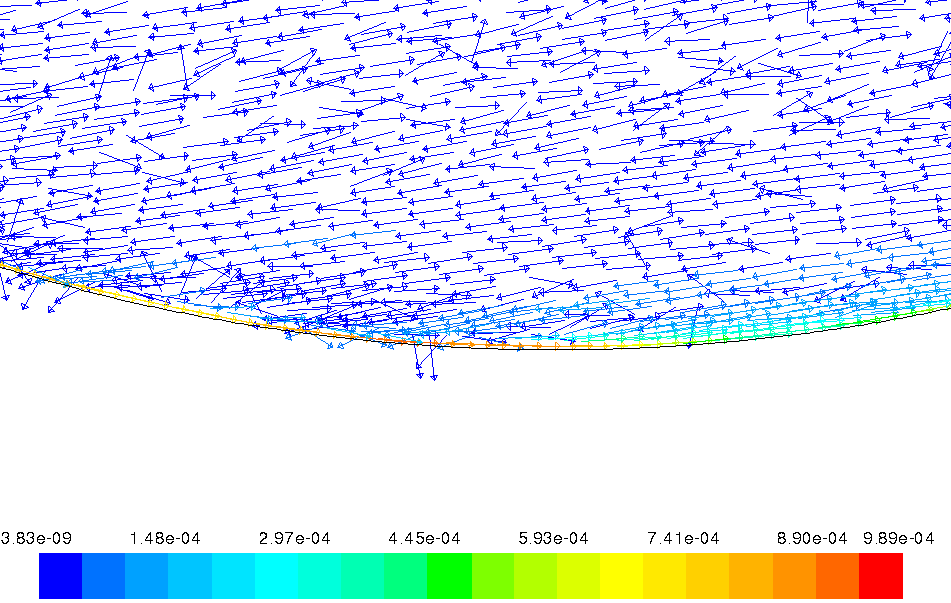} 
\caption{Strong currents along the IV boundary (in m/s). This feature is prominent and common to all IV models, except the "control", stably-stratified one with adiabatic boundary conditions. This shows the feature's model independence. Vectors show fluid movement direction in each of the modeled cells. Their length is normalized and should not be taken as a measure of the flow speed (hence spurious vectors pointing out of the main flow, or superimposed vectors with different directions, are expected as small vortices may appear, but are generally negligible considering their color-coded small magnitude and localization): this figure solely intends to illustrate the strong surface current, and its reduced radial extent. Color version available online.}
\label{fig:IV_lower_current} 
\end{figure}

Furthermore, a common characteristic to all models is a strong (up to 5 orders of magnitude above typical bulk currents) surface current along the vessel's internal periphery (see Figure~\ref{fig:IV_lower_current}), which in general does not span the whole surface with the same direction, owing to the inhomogeneous vertical separation between isotherms. It is worthwhile to note that this feature was not present in any area of the stably-stratified "control" model, indicating it is a feature uniquely derived from the boundary conditions and geometry, but definitely not numerically-induced. This feature was hypothesized as an explanation for the introduction of background components (particularly $^{210}$Po) from the less radio-pure vessel nylon into the FV's scintillator, although its detachment mechanism was the subject of much speculation and remained unexplained before numerical simulations were carried out.

Interestingly, this rules out the idea of a "chimney" effect picking up radio-impurities from the vessel and propelling them up or down the vertical axis of the vessel. For similar temperature fields, and changes, as the ones studied here for the 2015-17 timespan, the dominant mixing mechanism is instead mainly horizontal. Extended discussion of backgrounds stability and their transport with fluid movement will not be further discussed here, being beyond the scope of the current CFD work. This is mainly because of the status of the analysis techniques for reliably and unbiasedly counting the very low $^{210}$Po concentrations, regionally, throughout the IV ($\approx$20 counts/day/100 tonnes), and our capability to quantitatively relate that counting rate to the simulated fluid-dynamic profiles. Instead, an upcoming publication will report in detail the techniques used for low-statistics background tracking through data selection and localization, and its correlation with CFD scenarios under development.

Bulk horizontal velocities between $\mathcal{O}$(10$^{-5}$) and $\mathcal{O}$(10$^{-7}$) m/s were seen to be the largest in magnitude in all models, including the control scenario with ideally-stratified bulk with adiabatic boundary conditions. Even though these currents are seen to be mesh-enhanced locally, thanks to the aforementioned control stratified adiabatic case, large-scale features are thought to be physical: a markedly different horizontal current distribution is seen in the realistic scenarios, and much less horizontal organization develops in the stratified case, as well as with slower currents.

No large-scale organized vertical motion was seen to exist, and areas of vertical velocity larger than mesh cell hotspots (still, no larger than a few tens of centimeters in size) are typically between $\mathcal{O}$(10$^{-7}$) and $\mathcal{O}$(10$^{-9}$) m/s. This excludes the boundary layer in close proximity to the vessel boundary, where current magnitudes are much larger (up to $\approx$10$^{-3}$ m/s in some cases).

\section{Conclusions and prospects}
\label{sec:conclusions}

Borexino's sensitivity to solar neutrinos is determined by its unprecedentedly low intrinsic background levels, which in some cases can be indistinguishable from the Compton-scattered neutrino signal.  In particular, the CNO component is very sensitive to the $^{210}$Bi levels, whose $\beta$-decay signal is in principle indistinguishable from the CNO $\nu$s scattering and overlaps with it in the spectrum's region of interest. The $\alpha$-decaying daughter $^{210}$Po has shown temperature-correlated oscillations in its regional equilibrium rate, but would be the best handle to determine $^{210}$Bi levels with high precision if it were stable. In this paper, we have discussed the deployment of the global Borexino Thermal Monitoring and Management System (BTMMS), composed by the precision monitoring hardware (Latitudinal Temperature Probes System, LTPS) as well as the passive and active thermal management systems: Thermal Insulation System (TIS) and Active Gradient Stabilization System (AGSS), respectively. The BTMMS is aimed at increasing the detector's thermal stability with the objective of reducing scintillator mixing, exploiting the idea that background stability is directly influenced by fluid-dynamical stability. Ideally, this stabilization should reach a level which would not bring $^{210}$Po from the peripheral areas of the detector into the Fiducial Volume (FV). The deployment of the LTPS monitoring and TIS insulating systems have unequivocally shown that (i) the increase of the top-bottom positive stratification gradient in the fluids inside Borexino, with relatively small North-South asymmetries, and (ii) the smoothing of the external environment's thermal upsets that transmit toward Borexino's interior, have a direct and positive impact toward the stabilization and reduction of fluid mixing. Indeed, the stratification in Borexino's interior is interpreted as the most stable ever in the detector's life, separated by one of the greatest gradients ever achieved since filling.

Conductive Computational Fluid Dynamics (CFD) simulations widened the level of insight into the system's ideal, stable condition by allowing for a full detector picture, anchored on the empirical data the LTPS probes provide. In particular, excessive cooling of the full detector has been shown not to be of concern in the foreseeable future due to the TIS installation, given the contact with the ambient air will provide enough heat exchange to stabilize the top-bottom gradient once the bottom has cooled off. In fact, the TIS is confirmed to be greatly beneficial in order to limit and delay in time the conductive heat exchange between ambient air and the WT's water, which limits the amplitude of potential thermal oscillations that could trigger convection and propagate inside the Inner Vessel (IV). The cooling process is seen to be mostly-conduction dominated. Moreover, the cooling constant obtained is in agreement with the preliminary calculations derived from first-principles analysis and extrapolation of empirical LTPS data, suggesting an upper limit of lost power through the base heat sink. Structures have been shown not to act as significant heat bridges that could induce large-scale perturbations into an otherwise still fluid. Operation of AGSS within reasonable limits is shown to allow for the heat application to be restricted to the area of interest, and furthermore be an adequate measure to "freeze" the top isotherms in place, avoiding unwanted seasonal temperature inversions in Borexino's dome, where the stratification is the weakest.

Fully-convective, full-detector CFD simulations are impractical even in 2D due to computing and timing limitations. Furthermore, radial segmentation in Borexino makes simulating all fluid movements in the detector unnecessary, since only the temperature boundary conditions onto the IV will impact fluid behavior therein -- and, with it, radioisotope transport from the periphery inwards. A further in-depth study is in preparation, detailing the implications the discovered fluid-dynamic mechanisms have on background migration, as well as the data analysis techniques employed for background identification, tracking and regionalization.

The current work provides valuable insights into the determination of natural convection dynamics in a closed system with a pseudo-stable stratification such as Borexino, showing excellent benchmarking reproducibility in the Rayleigh number ranges of 10$^5$-10$^7$, and appropriate large-scale reproducibility down to $\mathcal{O}$(10$^3$). Attainment of global, organized convective modes is seen to exist in cylindrical geometries. Furthermore, no threshold is found with respect to the $\Delta T$ characteristics triggering these modes, as long as this $\Delta T$ is applied on the lateral walls. Conversely, only local convection will appear, up to the vertical distance the isotherms will be displaced to, when applying the $\Delta T$ to just the terminal caps. In a spherical geometry such as the Borexino detector, good thermal reproducibility was reached by comparing the large historical dataset of recorded temperatures to the temperature field obtained in CFD runs at the same positions. 

This allowed for the study of the smallest unsegmented region of interest possible, the Active Volume of the detector, where the understanding of weak scintillator-mixing currents is critical to further improve the radioactive background levels, as mentioned. Horizontal movement caused by lateral asymmetrical imbalances in the boundary conditions is seen to be the main driving factor in bringing fluid from the periphery to the center of the volume, while no detector-wide, vertical forced convective fields are seen to develop. Observed horizontal currents are seen to be of the order and span that would be of concern for background transport ($> \mathcal{O}$(10$^{-7}$) m/s), but this is not so for vertical currents. Carrying capacity maps ("iso-stream-function plots") clearly mark the regions where most of the fluid motion occurs, offering a powerful tool to understanding past behavior in the detector. They should also allow to engineer temperature profiles foreseen to cause minimal mixing, in order to inform future AGSS directives aimed toward establishing an ultra-low level of scintillator mixing -- and, with it, unprecedented levels of radioactive backgrounds in Borexino's FV.

We believe the present results may not only shed light on the particular case of the fluid-dynamics inside Borexino, but also expand the limited modeling of non-turbulent fluid mixing in pseudo-steady-state closed systems near equilibrium, subject to small asymmetrical perturbations in their temperature field, such as liquid reservoirs (water, liquified gas, petroleum-derived, deep cryogenics...), which share equivalent geometries and conditions.

\section{Acknowledgements}

The Borexino program is made possible by funding from INFN (Italy), NSF (USA), BMBF, DFG, HGF and MPG
(Germany), RFBR (Grants 16-02-01026 A, 15-02-02117 A, 16-29-13014 ofi-m, 17-02-00305 A) (Russia), and NCN
Poland (Grant No. UMO-2013/10/E/ST2/00180). We acknowledge the generous hospitality and support of the Laboratori Nazionali del Gran Sasso (Italy). All numerical simulations in this study are performed in the HPC system of the interdepartmental laboratory 'CFDHub' of Politecnico di Milano.

\section*{References}
\footnotesize
\bibliography{BTMMS}
\normalsize
\end{document}